\documentclass[
  pre,
  twocolumn,
  twoside,
  byrevtex,
  superscriptaddress,
  floatfix,
  longbibliography,
]{revtex4-2}

\usepackage[realmainfile]{currfile}
\input{\currfilebase.settings}
\usepackage{amssymb}
\usepackage{amsmath}

\usepackage{tcolorbox}

\usepackage{url}
\PassOptionsToPackage{hyphens}{url}\usepackage{hyperref}
\makeatletter
\g@addto@macro{\UrlBreaks}{\UrlOrds}
\makeatother

\usepackage{color, colortbl}

\definecolor{Gray}{gray}{0.9}

\definecolor{goodblue}{RGB}{0, 91, 187}
\usepackage{hyperref}
\hypersetup{
  colorlinks=true,
  allcolors=goodblue,
  urlcolor=goodblue,
  citecolor=goodblue,
  pdfborder={0 0 0},
  breaklinks=true,
}

\usepackage[normalem]{ulem}

\makeatletter

\def\CT@@do@color{%
  \global\let\CT@do@color\relax
  \@tempdima\wd\z@
  \advance\@tempdima\@tempdimb
  \advance\@tempdima\@tempdimc
  \advance\@tempdimb\tabcolsep
  \advance\@tempdimc\tabcolsep
  \advance\@tempdima2\tabcolsep
  \kern-\@tempdimb
  \leaders\vrule
  \hskip\@tempdima\@plus  1fill
  \kern-\@tempdimc
  \hskip-\wd\z@ \@plus -1fill }
\makeatother

\usepackage{xcolor}

\definecolor{olivegreen}{rgb}{0.33333,.41961,0.18431}
\definecolor{forestgreen}{rgb}{0.13333,.5451,0.13333}
\definecolor{lightgrey}{rgb}{0.7,0.7,0.7}
\definecolor{verylightgrey}{rgb}{0.90,0.90,0.90}
\definecolor{veryverylightgrey}{rgb}{0.95,0.95,0.95}
\definecolor{grey}{rgb}{0.5,0.5,0.5}

\definecolor{headerblue}{HTML}{33367E}
\definecolor{unitednationsblue}{HTML}{4D88FF}

\definecolor{charcoal}{HTML}{36454F}
\definecolor{cinerous}{HTML}{98817B}
\definecolor{feldgrau}{HTML}{4D5D53}
\definecolor{glaucous}{HTML}{6082B6}
\definecolor{arsenic}{HTML}{3B444B}
\definecolor{xanadu}{HTML}{738678}

\definecolor{firebrick}{HTML}{B22222}
\definecolor{orangered}{HTML}{FF4500}
\definecolor{tomato}{HTML}{FF6347}

\definecolor{purpletaupe}{HTML}{3B444B}

\usepackage{graphicx,epsfig,verbatim,enumerate}

\usepackage{ifthen}

\usepackage{longtable}

\usepackage{mathtools}

\newboolean{twocolswitch}

\newcommand{\sindex}[1]{}
\newcommand{\nindex}[1]{}

\newcommand{\etal}{\textit{et al.}}
\newcommand{\www}[1]{\url{#1}}

\usepackage{lettrine}

\setboolean{twocolswitch}{true}

\begin{document}

\title{\protect
  An assessment of measuring local levels of 
homelessness  \\
through proxy social media signals


}

\author{
\firstname{Yoshi Meke}
\surname{Bird}
}
\email{yoshi.bird@uvm.edu}

\affiliation{
  Computational Story Lab,
  Vermont Complex Systems Center,
  MassMutual Center of Excellence for Complex Systems and Data Science,
  Vermont Advanced Computing Center,
  University of Vermont,
  Burlington, VT 05405.
  }

\author{
\firstname{Sarah E.}
\surname{Grobe}
}

\affiliation{
  University of Vermont,
  Burlington, VT 05405.
  }

\author{
\firstname{Michael V.}
\surname{Arnold}
}

\affiliation{
  Computational Story Lab,
  Vermont Complex Systems Center,
  MassMutual Center of Excellence for Complex Systems and Data Science,
  Vermont Advanced Computing Center,
  University of Vermont,
  Burlington, VT 05405.
  }

\author{
\firstname{Sean P.}
\surname{Rogers}
}

\affiliation{
  University of Vermont,
  Burlington, VT 05405.
  }

\author{
\firstname{Mikaela I.}
\surname{Fudolig}
}

\affiliation{
  Computational Story Lab,
  Vermont Complex Systems Center,
  MassMutual Center of Excellence for Complex Systems and Data Science,
  Vermont Advanced Computing Center,
  University of Vermont,
  Burlington, VT 05405.
  }

\author{
\firstname{Julia Witte}
\surname{Zimmerman}
}

\affiliation{
  Computational Story Lab,
  Vermont Complex Systems Center,
  MassMutual Center of Excellence for Complex Systems and Data Science,
  Vermont Advanced Computing Center,
  University of Vermont,
  Burlington, VT 05405.
  }

\author{
\firstname{Christopher M.}
\surname{Danforth}
}

\affiliation{
  Computational Story Lab,
  Vermont Complex Systems Center,
  MassMutual Center of Excellence for Complex Systems and Data Science,
  Vermont Advanced Computing Center,
  University of Vermont,
  Burlington, VT 05405.
  }

\affiliation{
  Department of Mathematics and Statistics,
  University of Vermont,
  Burlington, VT 05405.
}

\author{
\firstname{Peter Sheridan}
\surname{Dodds}
}
\email{peter.dodds@uvm.edu}

\affiliation{
  Computational Story Lab,
  Vermont Complex Systems Center,
  MassMutual Center of Excellence for Complex Systems and Data Science,
  Vermont Advanced Computing Center,
  University of Vermont,
  Burlington, VT 05405.
  }

\affiliation{
  Department of Computer Science,
  University of Vermont,
  Burlington, VT 05405.
}

\affiliation{
  Santa Fe Institute,
  1399 Hyde Park Rd,
  Santa Fe, NM 87501.
}

\date{\today}

\begin{abstract}
  \protect
  Although nearly 600,000 people experience homelessness in the United States every year, efforts to address this public health crisis are limited by the underperformance of standard methods to estimate localized and nationwide homelessness. Recent studies suggest social media activity can function as a proxy for measures of state-level public health, detectable through straightforward applications of natural language processing. 
We present results of our efforts to apply this approach to estimate homelessness at the state level throughout the US during the period 2010-2019 and 2022 using a dataset of roughly 1 million geotagged tweets containing the substring ``homeless.''  Correlations between homelessness-related tweet counts and ranked per capita homelessness volume, but not general-population densities, suggest a relationship between the likelihood of Twitter users to personally encounter or observe homelessness in their everyday lives and their likelihood to communicate about it online. 
An increase to the log-odds of the word ``homeless'' appearing in an English-language tweet, as well as an acceleration in the increase in average tweet sentiment, suggest that tweets about homelessness are also affected by trends at the nation-scale. Additionally, changes to the lexical content of tweets over time suggest that reversals to the polarity of national or state-level trends may be detectable through an increase in political or service-sector language over the semantics of charity or direct appeals. Although tweet sentiment does not correlate to changes in homelessness volume, an analysis of user account type undertaken to explain nationwide sentiment dynamics revealed changes to Twitter-use patterns by accounts authored by individuals versus entities that may provide an additional signal to confirm changes to homelessness density in a given jurisdiction. While a computational approach to social media analysis may provide a low-cost, real-time dataset rich with information about nationwide and localized impacts of homelessness and homelessness policy, we find that practical issues abound, limiting the potential of social media as a proxy to complement other measures of homelessness.  
\end{abstract}



\maketitle


\section{Introduction}
\label{sec:measuring-homelessness.introdution}

In the United States---the world's wealthiest nation---nearly 600,000 individuals experience at least one night of homelessness every year~\cite{pitestimates}. One of the most critical social determinants of health, housing insecurity often prevents access to health-supporting resources including medical and sanitary services while at the same time increasing mental and physical stressors correlated with poor health outcomes, including higher rates of illness and death. 
In the US, a person experiencing homelessness is up to four times more likely to die prematurely and has a life expectancy of only 48 years~\cite{hchmd}. 

For the past five years, official counts of homelessness in the US have been on the rise after years of sustained progress in decreasing the raw count of homeless households nationwide~\cite{pitestimates}. 
It is now more important than ever that we develop cost-effective, high-impact, rapid-turnaround interventions that are sustainable, both from a collective resources perspective and from the perspective of households moving through the service delivery system.

Efforts to address homelessness, however, have long been troubled by our collective underperformance in measuring the extent of the problem and identifying appropriate geographic allocation of resources and interventions. Since the 1970s, federal policy bodies such as the US Department of Housing and Urban Development (HUD), the US Department of Agriculture (USDA), and the US Census Bureau, have attempted to develop and refine efforts to count the homeless with the stated policy goal of making homelessness 
``rare, brief, and non-recurring''~\cite{openingdoors}.  
The Point in Time (PiT) count, a national effort conducted and reported annually since 2007 by regional US service coordinating entries called Continua of Care (CoCs) in partnership with outreach volunteers, 
has been the federal standard for estimation of homelessness distribution~\cite{CoCcounty}.

The Point in Time Count, however, has come under increasing scrutiny, criticized for a lack of uniformity in data collection and estimation techniques, poor design, 
coarse geographic scales, 
and coarse time scales, 
which make meaningful research and analysis nearly impossible~\cite{wapo, CoCcounty}. 
In response, advocates within the homelessness services world increasingly call for new methods to generate more reliable, finely-grained estimates of homelessness distributions and system flow dynamics. Accordingly, in our present study, we aim to determine whether social media analytics could serve as a comparable point-in-time and/or dynamic proxy measurement of homelessness at the state level that could provide a finer temporal resolution than the PiT's annual timeframe. 

\subsection{Literature Review}
\label{sec:measuring-homelessness.litreview}

Recent research on homelessness and social media use has applied sentiment and content analysis to posts on homelessness from homeless versus non-homeless bloggers. 
In ``Characterizing Homeless Discourse on Social Media,'' researchers reviewed Tumblr posts with hashtags \texttt{\#homeless}, 
\texttt{\#homelessness}, 
\texttt{\#poverty}, 
\texttt{\#begging}, 
and \texttt{\#homelessshelter}, 
dividing users and posts into homeless versus non-homeless users and posts. Using word2vec embeddings and Latent Dirichlet Allocation (LDA) algorithms to cluster posts by semantic similarity, the authors found that content by homeless users had a more negative sentiment, used more first-person pronouns, and tended to chronicle the challenges they were experiencing in homelessness. By contrast, relevant posts by non-homeless users tended to use more abstract, depersonalized words about advocacy, acts of kindness, or the news~\cite{clusterlitreview}.

Other research has focused on user-role identification and network analysis of homeless users' social media accounts. 
For example, in Koepfler \etal's analysis of @WeAreVisible, a Twitter account encouraging content visibility by and for people experiencing homelessness, identified eleven primary user roles and analyzed the structure of the connections between and within user types~\cite{wearevisiblefull}. 
The majority of homeless group participants belonged to a densely-connected cluster of users comprising around 10\% of all members. The authors found that, despite the community's stated purpose, only about 4\% of users in the account's network self-identified as homeless, with over half the network being tagged as ``social media enthusiasts'' 
or ``do-gooders''~\footnote{
    Follow-up work moved away from account labels to account ``associations'' (i.e., characteristics) in acknowledgement that individual users may have more than one account and/or that classifications were not necessarily mutually exclusive---for example, some ``non-profit generalists'' had prior lived experience of homelessness.}.
People experiencing homelessness tended to be well-connected with one another and connected, though to a lesser extent, with other user types \cite{wearevisibleshort}. Koepfler \etal\ also sought to distinguish values characterizing homeless versus non-homeless users on the basis of their tweet content, finding that homeless and formerly-homeless users tended to express, 
among other values, 
helpfulness, 
broadmindedness, 
justice, 
equality, 
responsibility, 
and freedom more often than never-homeless users.

Our concern is both broadly the analysis of volume, sentiment, and content of homelessness-related Twitter use, and also narrowly its relationship to observed measures of homelessness rates within a US-geotagged jurisdiction. Prior research has established correlations between tweet content, location, and sentiment with other public health metrics. For example, co-authors developed the ``Lexicocalorimeter'', 
a tool ranking states by caloric intake and output estimates assigned to US-geotagged tweets containing food, drink, and activity names. Through comparison between state-level tweet content, a range of strong, common-sense relationships emerged between aggregate caloric scores and an assortment of health and well-being indicators, such as life expectancy, obesity rates, and mental health challenges \cite{lexicocalorimeter}.  Similarly, researchers have demonstrated strong correlations between state-level expressions of happiness, measured via tweet sentiment, and well-established but resource-intensive survey measures of happiness like the Gallup well-being index \cite{happiness}. Importantly, research has also suggested the capacity of social media sentiment and content analysis to serve as a real-time public health surveillance tool through geolocation analysis of tweets containing symptom-descriptive words, a use-case of which was the monitoring of intra-urban spread of dengue fever in Indonesia \cite{denguefeverI, denguefeverII}. Thus, we were motivated to perform similar analysis of Twitter data to determine whether or not the platform could be harnessed to solve the seemingly-intractable problem of estimating homelessness---both comparative point-in-time levels across states and national and state-level trends over time.

\section{Data}
\label{sec:measuring-homelessness.data}

\textbf{Tweets:} 
Using the Twitter API v 2.0, which granted researchers full-archive access to, among other data, tweet and user account information, we queried all English-language tweets containing the word ``homeless'' that also contained geotag bounding box and other metadata indicating that they had originated from a location within the United States between the dates March 1, 2010, and December 31, 2022~\footnote{No tweets were returned for the period January 1, 2010, through February 28, 2010.}. 
The fields we accessed included:

\begin{itemize}
    \item 
    \textit{id:} A unique numeric identifier for each tweet.
    \item 
    \textit{text:} A string of words, numbers, characters, and emojis comprising the published content.
    \item 
    \textit{created\_at:} Date and time stamp of the tweet's publication.
    \item 
    \textit{author\_id:} A unique numeric identifier for the user who posted the tweet.
    \item 
    \textit{author:} A dictionary of descriptive attributes of the user, including user-generated, self-descriptive text and the username.
    \item 
    \textit{geo:} A dictionary of geolocation information, including country of origin, bounding box latitude/longitude coordinates surrounding the user's exact location at time of posting, a ``full\_name'' string descriptor of the location, and the place type corresponding to the named location (e.g., ``city'').
\end{itemize}

\begin{figure}[t!]
    \centering
    \includegraphics[width=\columnwidth]{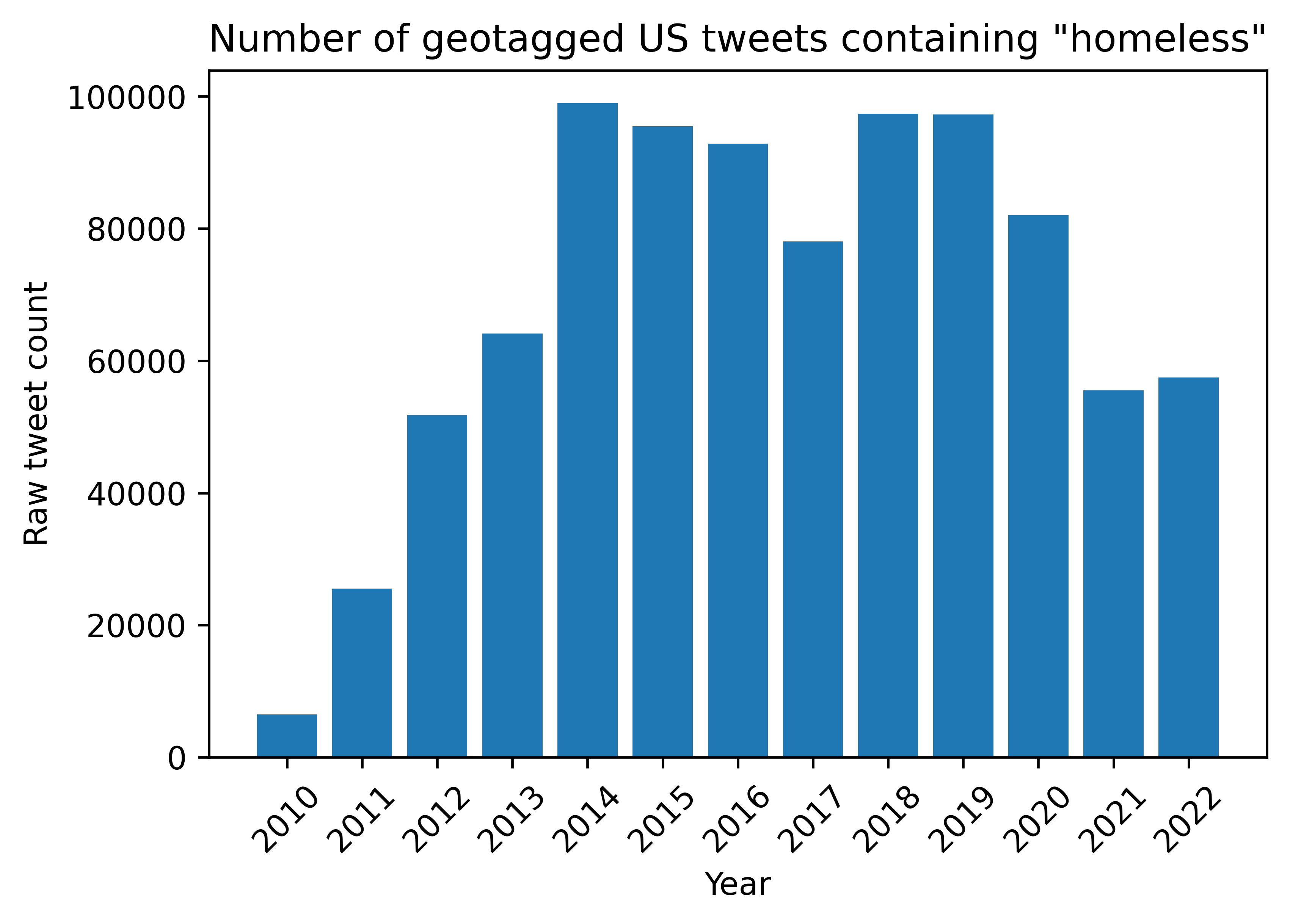}
    \caption{\textbf{Raw counts of all US-geotagged tweets containing ``homeless'' by year.} Geolocation functionality first became available in November 2009 but was only slowly rolled out through first months of 2010 to broaden its potential scope to all users and tweets, resulting in lower counts during the early years of available data.}
    \label{fig:tweet_histogram}
\end{figure}

We supplemented our dataset using the Storywrangler platform~\cite{storywrangler}, accessing log-odds ratio data for words (1-grams) across a random sample of 10\% of all English-language tweets during the same time period. Log-odds counts the number of English-language 1-grams containing the substring ``homeless'' as a fraction of all English-language 1-grams, transformed using the base-10 logarithm for clearer comparison across time steps. Data accessed by the Storywrangler site was made available to the University of Vermont via Gnip's Decahose application programming interface (API), an enterprise endpoint and streaming connection to Twitter data that has greatly enabled social media research and statistical analysis on a representative and robust sample of tweets.

\textbf{Homelessness data:} 
From \url{data.hud.gov}, we downloaded the 2007-2019 and 2022 Point in Time Estimates by CoC (Continuum of Care), a collection of annual reported counts of homeless individuals and families across the US reflecting a single 24-hour window in the last week of 
January~\footnote{Due to many states' departures from previous Point-in-Time count practices during the COVID-19 pandemic, the 2020 and 2021 counts were excluded from our analysis.}. 
Over time, the PiT dataset has evolved from just 28 questions in 2008 to 574 in 2022. In order to compare data across years, we selected the variable ``overall homeless'', an aggregate count of all sheltered and unsheltered adults and children identified as homeless. Initially, we hypothesized that sentiment and tweet volume would correlate most strongly with unsheltered homelessness, the visibility of which can trigger strong emotions. However, unsheltered homeless counts were characterized by much narrower variance, which could make correlations harder to discern, and relationships appeared stronger between total homelessness and other variables. Accordingly, our results focus exclusively on the overall homeless variable common to all years of the PiT count dataset.

\textbf{Census data:} To estimate annual per capita rates of homelessness and Twitter activity, we used the US Census Bureau's American Communities Survey One-Year Annual Estimates \cite{popestimates}. Within this dataset, we used only total state population estimates within each given year. To estimate the population density of homelessness (number of persons experiencing homelessness per square land-mile), we used the Land Area estimates for each US state as published by the Census Bureau in its State Area Measurements and Internal Point Coordinates dataset \cite{landestimates}. 

\subsection{Pre-processing}

Our dataset of geotagged US tweets contained 923,385 messages posted from March 1, 2010, through December 31, 2022, featuring a ``geoid'' field comprised of a dictionary of information about the specific location from which the tweet was published. We used information provided in the `geo' field to identify the state where the tweet was originally 
posted~\footnote{
We did not consider the impact of proxy servers on location distribution.}, and where the value was ambiguous (e.g., Starbucks), we applied GeoPy's Nominatim object and geolocator.reverse() function to the latitude and longitude coordinates, if available \cite{geopy}. Among our tweets, only 19,461 were unable to be precisely state-labeled, or 2.1\% of the dataset. We then added several fields, including ``state-year,'' for ease of filtering. Tweets were included in our analysis if they were geolocated to one of the fifty US states. We excluded tweets from the District of Columbia and US territories such as Guam or Puerto Rico.

For tweet text analysis, we next removed from the ``text'' field emoji, URLs, usernames (as indicated by an asperand symbol), and stop words that add no semantic meaning and, due to their frequency, dilute composite sentiment toward a neutral score (e.g., ``the'' and ``of''), retaining only blocks of alphanumeric characters separated by spaces into single-word tokens. 
Time stamps were converted from the API's Coordinated Universal Time (UTC) measurement to the local time zone at point of tweeting.
 Additionally, any hashtags in camel case were separated into their individual words (e.g., the hashtag \texttt{\#HomelessnessFirst} would be converted into the string ``homelessness first'' while \texttt{\#Homelessnessfirst} would not be affected).

\begin{figure*}
    \centering
    \includegraphics[width=\linewidth]{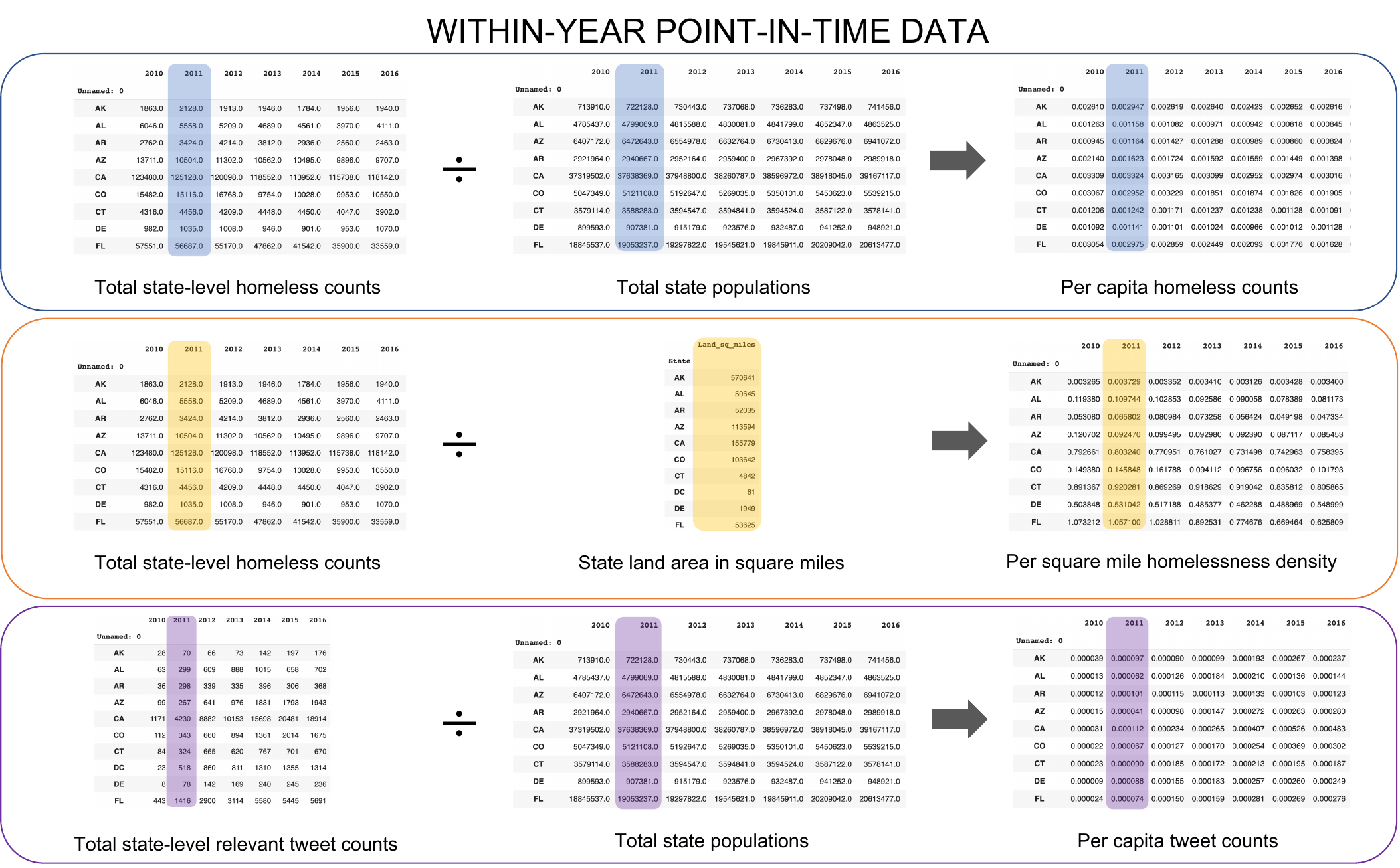}
    \caption{
    \textbf{Computation of point-in-time cross-state comparative data.} To re-scale measurements across states of different size, we calculated each state's homelessness counts or tweet counts by some scaling factor: each state's estimated population in the relevant year (top and bottom), estimated land area in square miles (center). Within-year correlation coefficients show us that states with highly-ranked per capita and per square land mile homelessness counts tend to have highly-ranked per capita homelessness-related tweet counts. The re-scaling helps us compare large and small states with the same statistical test such that state size does not account for observed correlations. See Fig.~\ref{fig:measuring-homelessness.glossary} for more details.}
    \label{fig:vertical}
\end{figure*}

\begin{figure*}
    \centering
    \includegraphics[width=\linewidth]{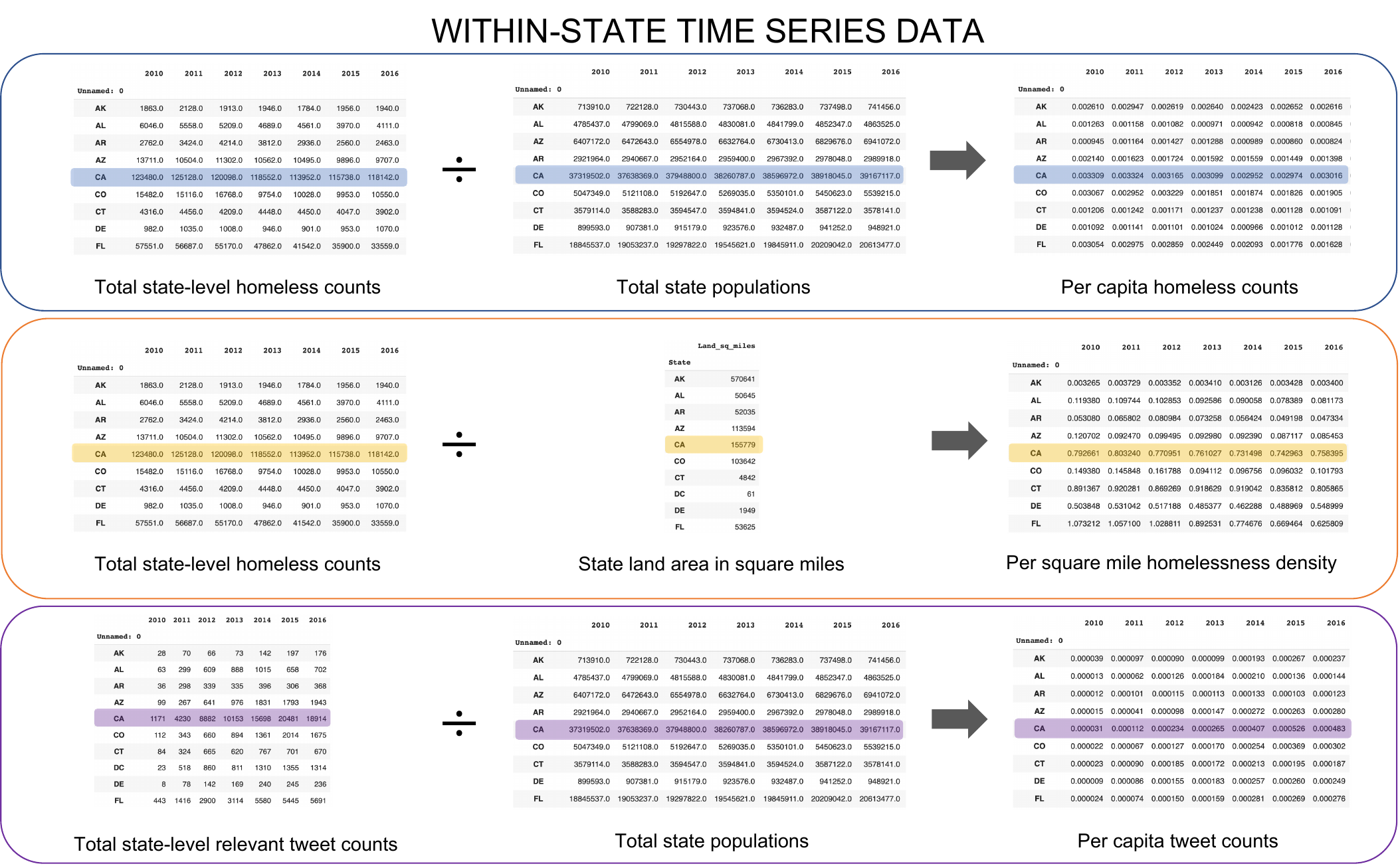}
    \caption{
    \textbf{Computation of within-state time series data.} We performed re-scaling as described at Fig.~\ref{fig:vertical} and then performed correlation tests to determine whether, within a single state's ten years of data, higher-ranked homeless counts resulted in higher-ranked tweet counts. This might suggest that Twitter users were aware of, and responding to, larger or smaller homeless counts within their state from year to year.}
    \label{fig:horizontal_calc}
\end{figure*}

Finally, we processed each dataset by transforming each state's raw counts to per capita counts and per square mile densities based on the annual census population estimate for each state (see Figures \ref{fig:vertical} and \ref{fig:horizontal_calc}). We then calculated annualized changes to the per capita tweet and homelessness count by state-year and log-transformed each as appropriate.

\section{Methods}

In our research, we consider the potential of real-time analysis of volume, content, and sentiment of homelessness-related social media communication to serve as a proxy variable estimating local homelessness rates or trends at the state level in the US. We initially hypothesized that localized changes in homelessness rates would exhibit a positive association with the volume of tweets generated within a given state, while sentiment expressed in tweets would be negatively correlated with changes in homelessness rates. Our proposed mechanism for these associations was that an increase in real-life, public encounters between the general public and individuals experiencing homelessness would result in an increase in online expressions of frustration and stigma, either against the structures that cause homelessness or toward the homeless themselves. However, social media users' abilities to perceive changes to homelessness are heterogeneous, highly variable, and context-dependent, unlike more ubiquitous phenomena previously explored, such as the sleep loss insult of daylight savings time \cite{sleeploss}, happiness \cite{happiness}, or caloric intake/output \cite{lexicocalorimeter}. Accordingly, we needed to consider a range of possible variables, measures, and relationships.

Testing for parametric ($r$) and nonparametric rank-rank ($\rho$) correlations (see Sec.~\ref{sec:correlating}), as well as time series cross-correlations (see Sec.~\ref{sec: crosscorrelations}), we analyzed relationships between and among:

\begin{itemize}
\itemsep0em 
    \item{Reported counts of homeless individuals,}
    \item{Tweet volume,}
    \item{Tweet sentiment, and}
    \item{Tweet content}
\end{itemize}

Each variable was measured:

    \begin{itemize}
    \itemsep0em 
    \item{As raw, per capita, and ranked counts, per square land-mile densities, and annualized changes thereof,}
    \item{As percentages of total nationwide measurement in a given year, and}
    \item{Aggregated by year across all states, by state across all years, and by state by year (``state-year''), }
    \item{With each measure additionally log-transformed as useful or appropriate.}
    \end{itemize}

We also compared time series data illustrating the log-odds of the substring ``homeless'' appearing in an English-language tweet with  nationwide homeless counts (see Sec.~\ref{sec:logodds}). 

Finally, we drew on techniques specific to the domain of natural language processing---sentiment analysis and allotaxonometry\textemdash to better understand the dynamics of sentiment and content expressed in homelessness-related tweets over time and at various scales (see Sections \ref{sec:sentiment} and \ref{sec:allotax}). 

\section{Results}

\subsection{Homelessness \& Tweet Volume}
\label{sec:correlating}
\subsubsection{Homelessness volume through time: Counts \& densities}

Although homelessness counts nationwide have been on the rise since 2017 following nearly a decade of annual decreases, mean and median per capita homelessness rates by state, as well as the maximum per capita rate nationwide, actually declined from 2012-2019 (see Figure 2).  
Moreover, while homelessness counts are characterized by a consistent annual variance of 0.000001 people experiencing homelessness per capita, the variance of estimated homelessness density is much higher, with a mean variance across states of 0.24 people experiencing homelessness per square land-mile. 
Nevertheless, there is a much higher rank-turbulence~\cite{allotaxonometry} among per capita state homelessness counts than among per square mile densities, with only two states consistently ranked in the bottom ten states for per capita counts and all ten of the bottom ten states for density remaining the same across all years of the data. 
Likewise, eight of the top ten density states were invariant across all years, as compared with only six of the top ten per capita states. 
The annualized distribution of raw state-level tweet counts had a variance distribution among all fifty states for each year of $3.06$$\times$$10^{4}$ to $1.22$$\times$$10^{7}$ per year, with state population predictably correlating to a state's raw tweet count across all years.

\begin{figure}
    \centering
    \includegraphics[width=\columnwidth]{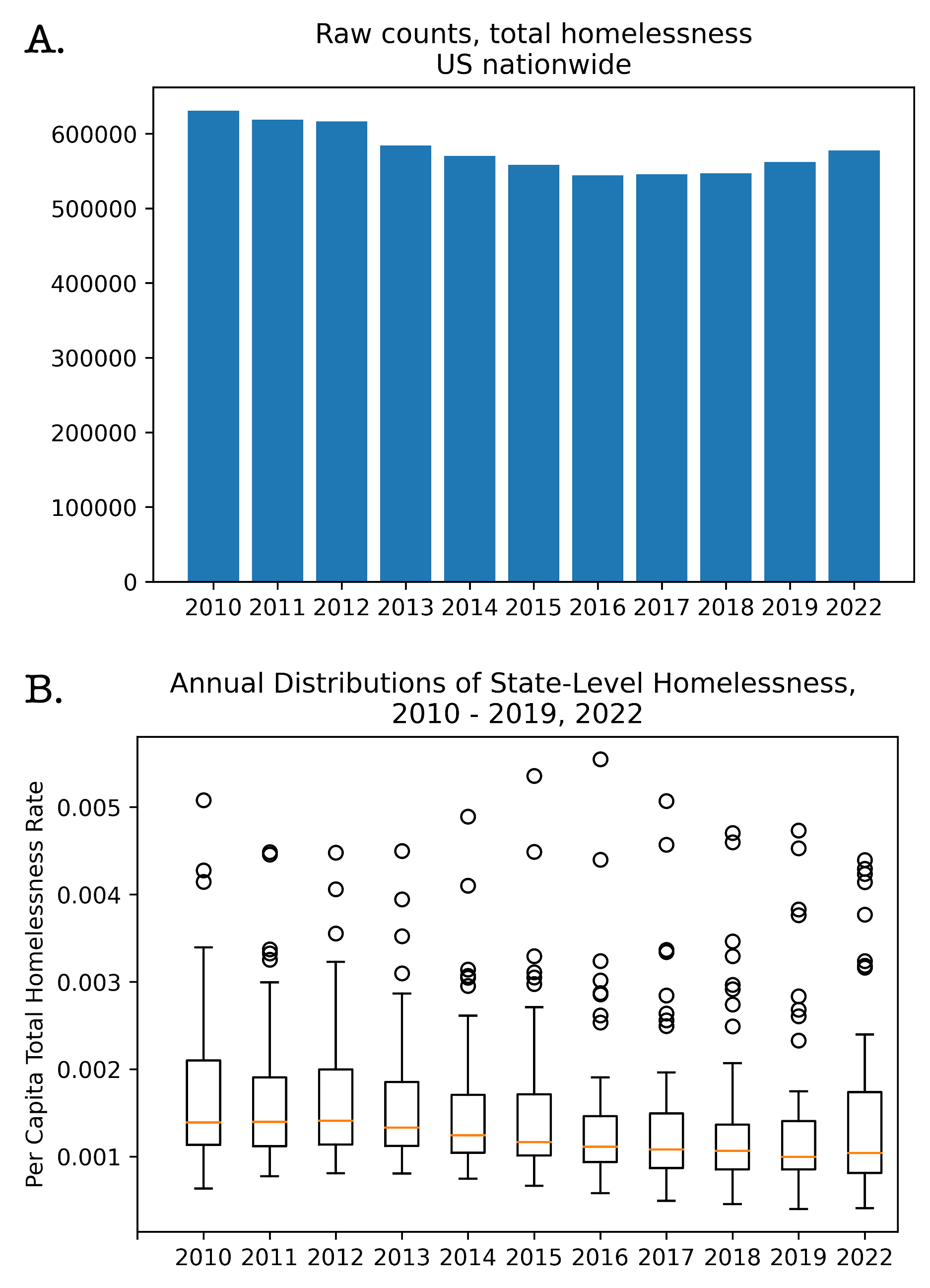}
    \caption{\textbf{A. Raw nationwide counts and B. per capita state homelessness rate distributions, as measured in the 2010-2019 and 2022 PiTs.} Nationwide US counts of total homelessness are on the rise since the 2017 Point-in-Time Count. 
    Roughly 75\% of the per capita counts for all fifty US states in each year falls below 0.002 people experiencing homelessness per capita (or 2 for every 1,000), though a heavy-tailed distribution across all years reveals a number of states in every year experiencing a per capita count at least twice as high.}
    \label{fig:raw_tweet_counts}
\end{figure}

\subsubsection{Correlating homelessness density, counts with tweet counts}

We found that the correlation between a state's ranked homelessness density and its ranked per capita homelessness-related Twitter activity is statistically significant across all years of data after 2010 (see Tab.~\ref{tab:rates}). We also observed that the ranked per capita homelessness counts correlates with ranked per capita homelessness-related tweet counts among states from 2013-2019 and again in 2022. 
We considered the possibility that state population density in general---and not the density only or specifically of people experiencing homelessness---was correlated with per capita tweet rates without regard for topic, which could provide an alternative explanation for the statistical significance of the relationship between homelessness density and tweet counts. 
We did not find, however, that overall state general-population density (number of persons per square land-mile) correlated at statistically significant levels with homelessness-related Twitter activity volume in any year.  

\begin{table*}[htp]
    \scriptsize
    \centering
    \begin{tabular}{|c|c|c|c|c|c|c|c|c|c|c|c|}
    \multicolumn{12}{c}{\textbf{State-level general-population densities, per capita tweet rates}}\\
          \hline
          & 2010 & 2011 & 2012 & 2013 & 2014& 2015 & 2016 & 2017 & 2018 & 2019 & 2022 \\
          \hline
         Spearman $\rho$ & -0.071 & -0.019 & 0.103 & 0.066 & 0.054 & 0.083 & -0.064 & -0.097 & -0.015 & -0.064 & -0.143 \\
         $p$-value & 0.619 & 0.896 & 0.473 & 0.645 & 0.707 & 0.563 & 0.653 & 0.497 & 0.914 & 0.658 & 0.318\\
         \hline

    \multicolumn{12}{c}{\textbf{State-level homeless-population densities, per capita tweet rates}}\\
          \hline
          & 2010 & 2011 & 2012 & 2013 & 2014& 2015 & 2016 & 2017 & 2018 & 2019 & 2022 \\
          \hline
         Spearman $\rho$ & 0.060 & 0.672 & 0.822 & 0.669 & 0.689 & 0.634 & 0.638 & 0.613 & 0.685 & 0.52 & 0.452 \\
         $p$-value & 0.677 & 6.64e-8 & 1.46e-13 & 7.93e-8 & 2.13e-8 & 6.00e-7 & 4.66e-7 & 1.77e-6 & 2.87e-8 & 9.26e-5 & 8.59e-4\\
         \hline
    \end{tabular}
    \caption{
    \textbf{Spearman $\rho$ correlation between population density and homelessness-related tweet rates.} Nonparametric, ranked correlations were found to be statistically significant with respect to homelessness densities (bottom) but not general-population densities (top).
    }
    \label{tab:rates}
\end{table*}

We propose that the likelihood of an average social media user to tweet about homelessness is greater if the probability is comparatively high that they will directly observe changes to their state's overall homeless population. This probability will depend, in part, on the size and population density distribution of their state of residence and the extent to which a raw change in homelessness counts is visible as a proportion of nationwide count or per capita proportion of a single state's total population. 
For example, in California, a total homeless count increase of 21,306 between 2018 and 2019---which translates to a change of 0.00054 persons experiencing homelessness per capita and 0.137 per square mile---may be less noticeable than a decrease of just 1,597 in Massachusetts over the same period---a decrease of half as many persons per capita (0.00024) but nearly twice as many per square mile (0.20).

\subsubsection{Tracking nationwide counts, log-odds of homelessness-related tweeting}
\label{sec:logodds}

\begin{figure}
    \centering
    \includegraphics[width=\linewidth]{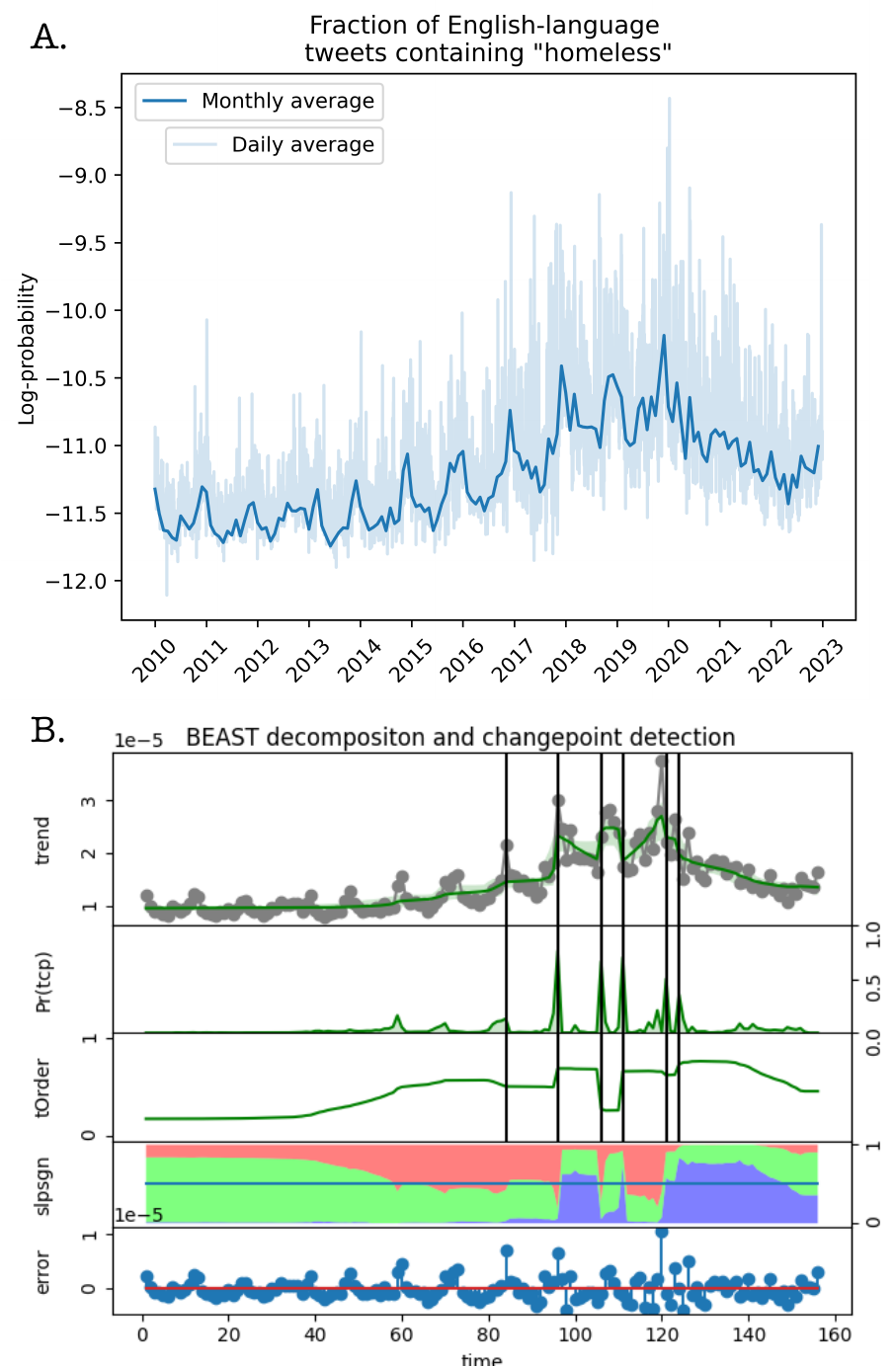}
    \caption{\textbf{A. Log-probability of the 1-gram ``homeless'' appearing in an English-language tweets and B. BEAST trend changepoint detection visualization for month-scale sentiment data.} Drawing from a random sample of 10\% of all English-language tweets, \url{storywrangler.org} provides statistics on the rank and frequency of $n$-grams, available at various timescales. The light blue line in the upper plot represents daily probabilities, while the dark blue represents monthly averages in the period 1/1/2010 - 12/31/2022. As with the tweet sentiment time series, the Rbeast package identifies time step 96 of the monthly average log-odds data, or January 2018, as a potential changepoint with an associated probability of 96.5\%.}
    \label{fig:odds}
\end{figure}

However, the salience of homelessness as a topic of interest for an average Twitter user is not only affected by the conditions in their local environment. National trends and the attendant media attention they receive, amplified by real-time distribution available on social media, may also affect the likelihood of the 1-gram ``homeless'' appearing in an English-language tweet. For example, as US trends shifted from decreasing to increasing nationwide counts, the log-odds of ``homeless'' appearing in a tweet, increasing on average $6.11\times 10^{-7}$ units per year (from $1.00\times 10^{-5}$ to $1.43\times 10^{-5}$ ) over the seven year period 2010-2016, increase at nearly six times that rate between 2017 and 2019 ($3.65\times 10^{-6}$ per year) \cite{storywrangler}, see Fig.~\ref{fig:odds}. This dramatic increase in proportion of homelessness-related tweets as a proportion of all Twitter content as US homelessness increased, paired with a similar acceleration in the increase to homelessness-related tweet positivity (see Sec.~\ref{sec:sentiment}), strongly suggest that homelessness-related Twitter activity is not only affected by the local density experienced by a given Twitter user, but also that user's contextualization of homelessness within broader national trends.  

\begin{figure}
    \centering
    \includegraphics[width=\columnwidth]{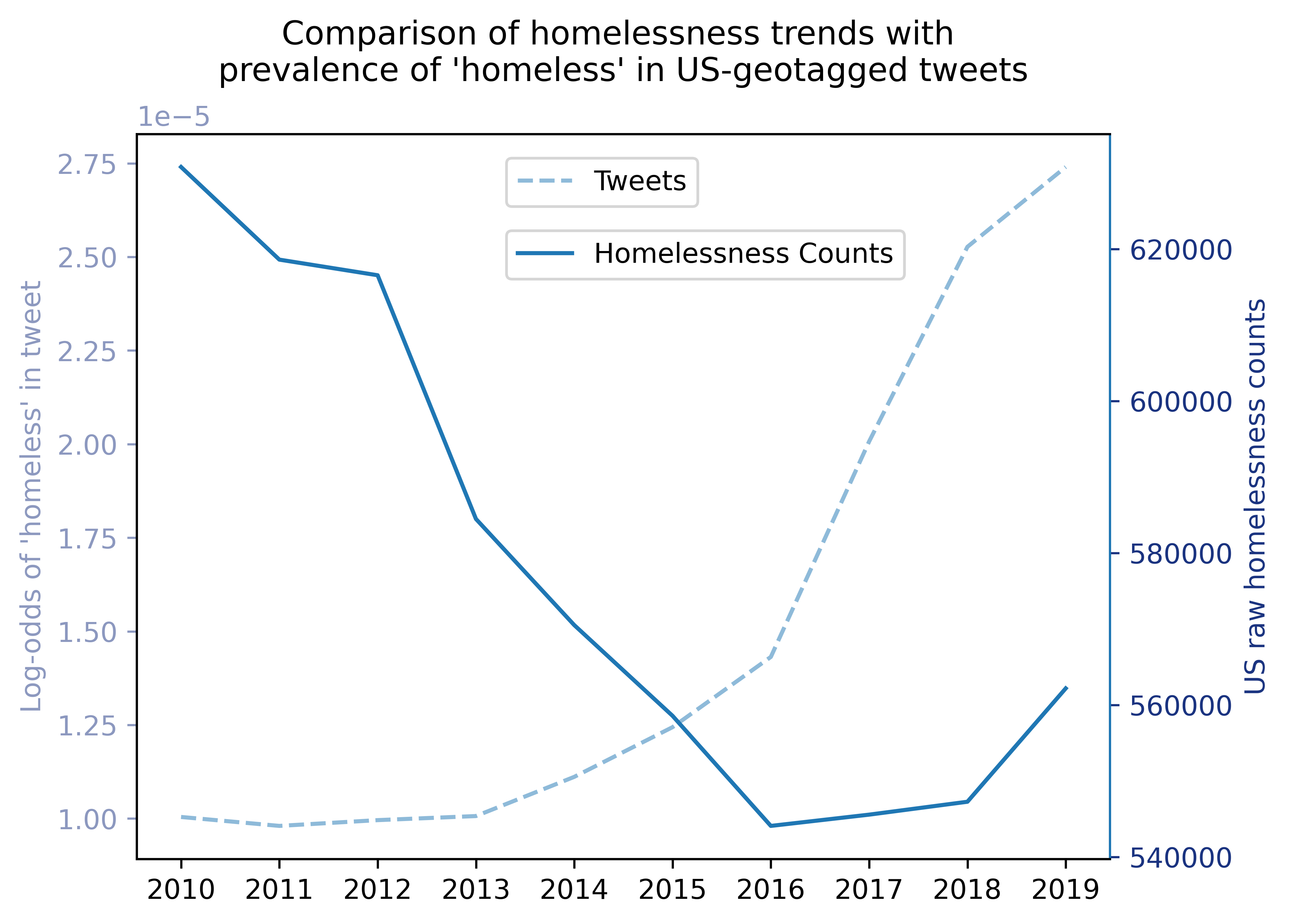}
    \caption{\textbf{Nationwide US homeless counts and log-odds of the 1-gram ``homeless'' appearing in an English-language tweet, annual scale.} Here, log-odds of the substring ``homeless'' appearing in a US-geotagged, English-language tweet were calculated for each month of the period 1/1/2010--12/31/2022 and are compared to nationwide raw homeless counts. Note the scaling of the double y-axes. See also \ref{Fig:unhoused}}
    \label{fig:logodds}
\end{figure}

\begin{figure*}
    \centering
    \includegraphics[width=\textwidth]{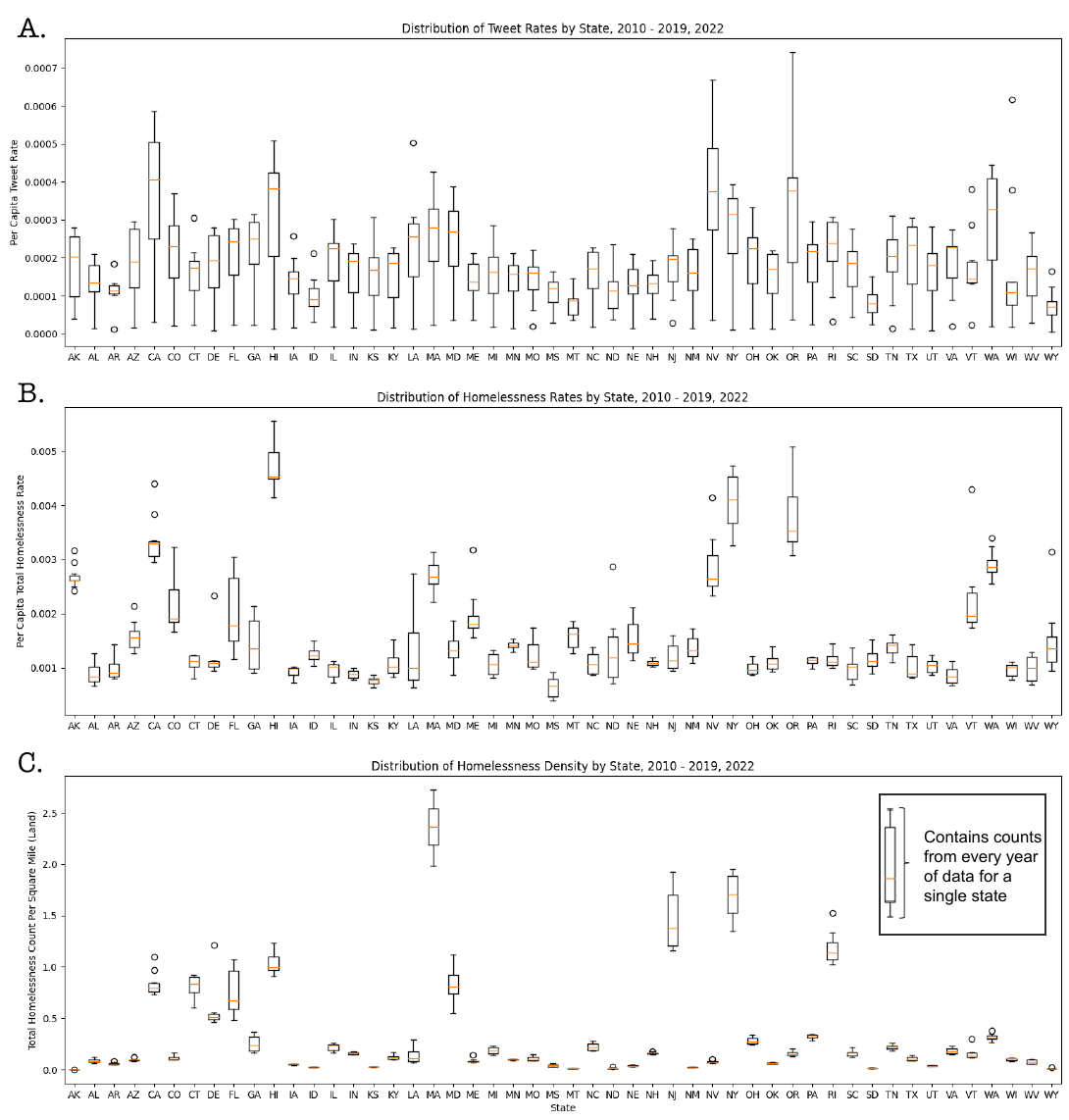}
    \caption{\textbf{Distribution of homelessness and tweet volume by state.} A. Distributions include per capita tweet rate, B. per capita homelessness rate, and C. per square land-mile homelessness density. Each box-and-whiskers plot displays the distribution of a single state's data across the years 2010-2019 and in 2022.}
    \label{fig:boxplotmania}
\end{figure*}

\begin{figure}
    \centering
    \includegraphics[width=\columnwidth]{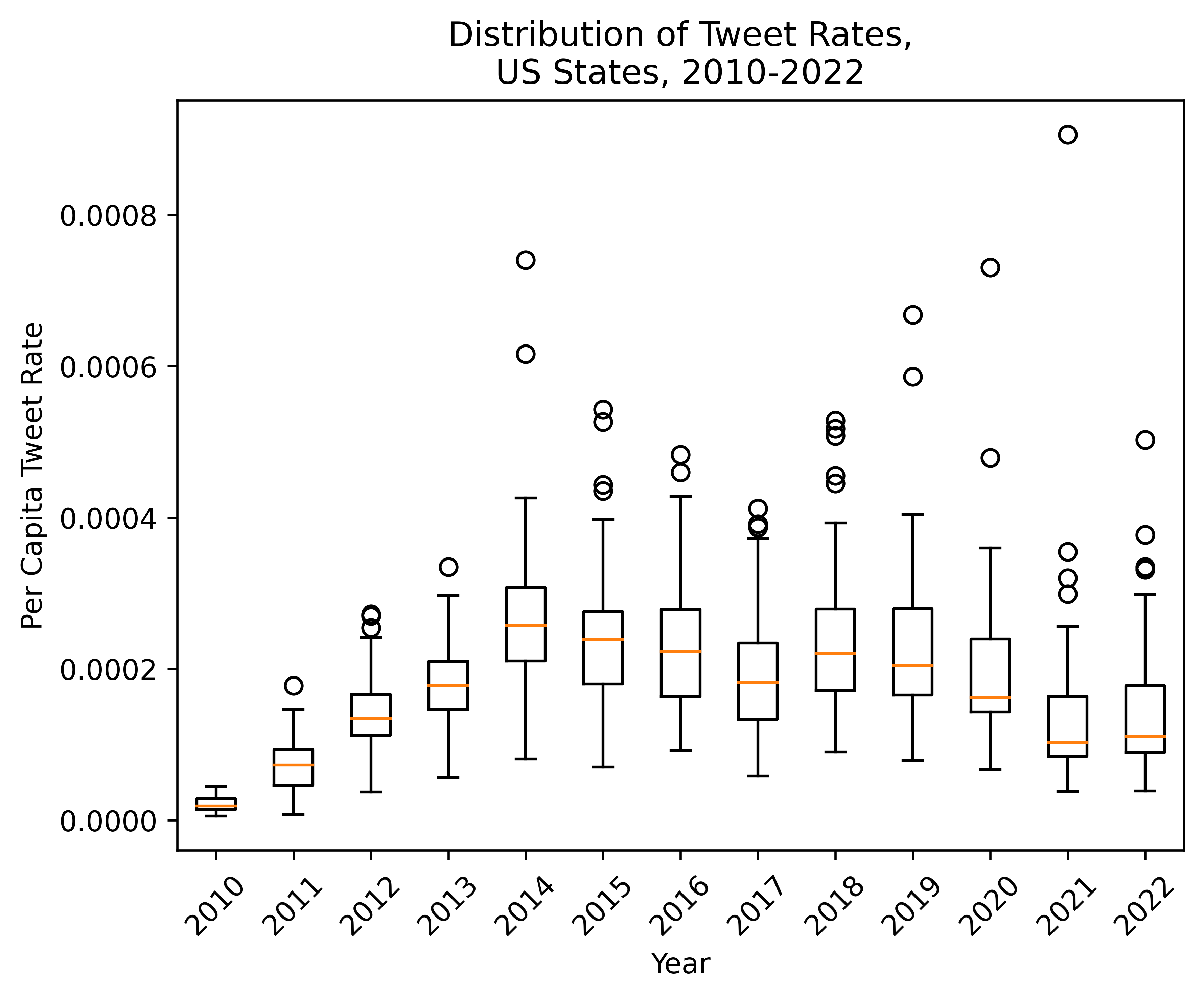}
    \caption{\textbf{Per capita tweet rate distributions by year, all US states.} Each box-and-whiskers plot represents the distribution of per capita tweet rates across all fifty US states in a single year.}
    \label{fig:tweet_state}
\end{figure}

\subsubsection{Within-state time series correlations}
\label{sec: crosscorrelations}

We next sought to determine whether fluctuations within a single state across time might correlate to changes in the per capita homelessness-related tweet count within that state. While in-year changes to per capita homelessness counts (see Fig.~\ref{fig:vertical}) did not correlate across states to per capita tweet counts, nineteen states exhibited statistically significant nonparametric rank-rank correlations between some measure of localized homelessness (count, density, or change since the prior year) and Twitter volume (count or change) across all eleven years of available data (see Fig.~\ref{fig:horizontal_calc}). Among these states, five (California, Florida, Delaware, Massachusetts, and New York) were both at or above the 75$^{\textnormal{th}}$ percentile of per capita homelessness densities across all ten years and also ranked in the top ten states with respect to variance in homelessness rate or density, indicating both high point-in-time measures of homelessness and longitudinal instability. An additional state (North Dakota) ranked within the top ten states with respect to variance only, likely because of low volume; one more (Washington) was consistently at or above the 75$^\textnormal{th}$ percentile of per capita homelessness rates. 

\subsubsection{Within-state time series cross-correlations}

We also examined whether time series data for states' homelessness counts exhibited cross-correlation with tweet counts after some period of lag, indicating for example that an increase in per capita homelessness would correlate to an increase in per capita tweets over the following $t$ years. Accordingly, we constructed two sets of time series data for each of the 50 states, then calculated the cross-correlation coefficients for up to 9 years of lag. Fig.~\ref{Fig:crosscorrboxplt} visualizes the distribution of cross-correlation measures across all states throughout the entire time period.

As expected, cross-correlations between point-in-time homelessness and tweet counts indicate that tweet volume likely reflects observations that are recent in time, i.e., within the last 1 to 2 years. The strongest cross-correlation relationships between per capita total homelessness and tweet counts were at time step $t=1$ and $t=4$, with Southern states Louisiana, Florida, South Carolina, and Virginia exhibiting a strong anti-correlation and New York and Massachusetts strong positive correlations 
(see Tab.~\ref{tab:crosscorrcoeff}, \ref{Fig:crosscorrboxplt}).
States with cross-correlation coefficients closest to 0 at $t=1$, on the other hand, included Arkansas, Pennsylvania, Maine, and California. 

\begin{figure*}
    \centering
    \includegraphics[width=6 in]{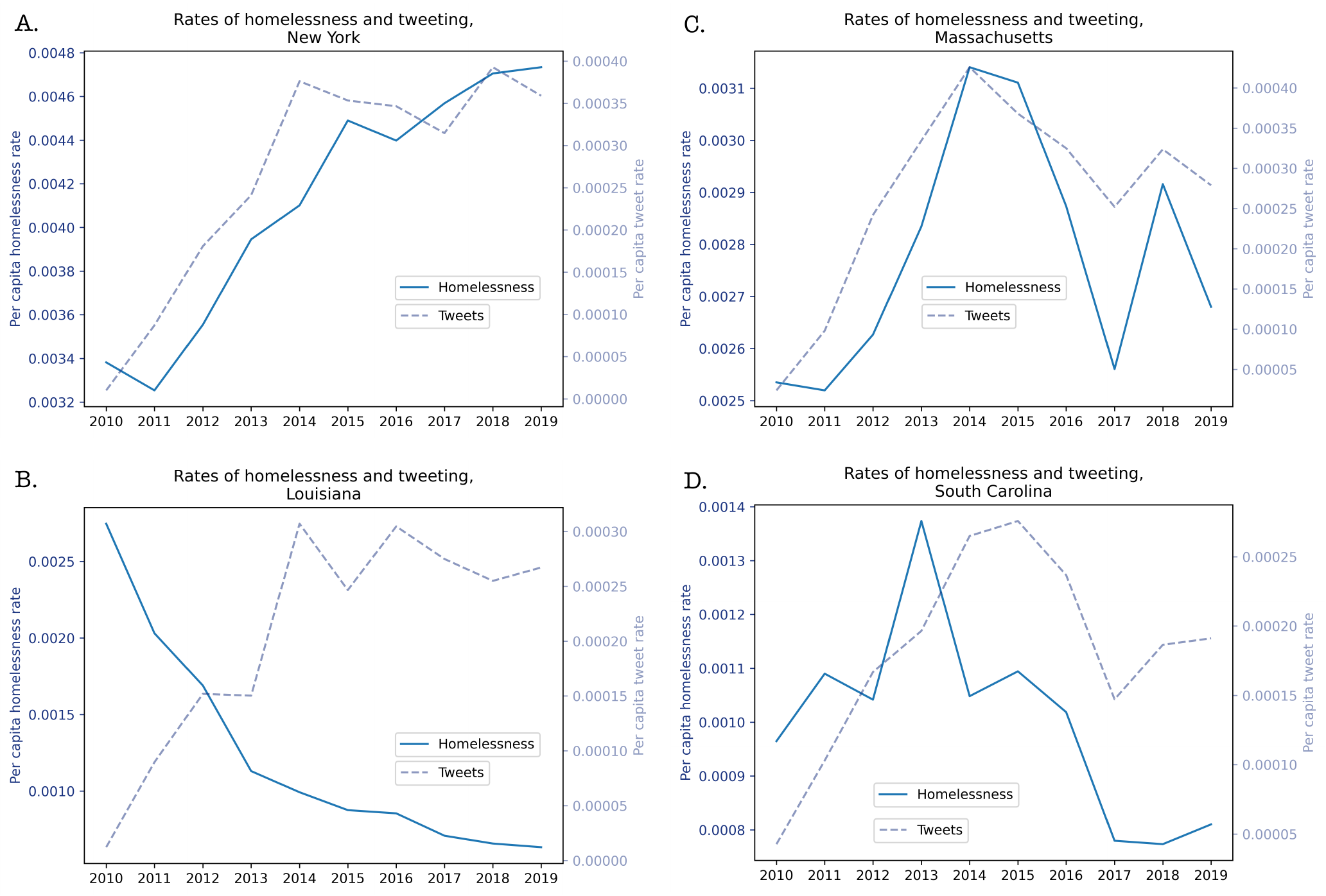}
    \caption{\textbf{Comparison of strongly correlated (A. New York, C. Massachusetts) and anti-correlated (B. Louisiana, D. South Carolina) time series of per capita homelessness and tweet rates.} Note that, while many Southern states see anti-cross-correlation between homelessness and tweet rates due to monotonically decreasing rates of homelessness offset by one year from primarily-increasing rates of tweeting, South Carolina's anti-correlation may be an artifact of the annual oscillation of its rates.}
    \label{fig:NYMALASC}
\end{figure*}

We also constructed time series for changes to per capita homelessness counts and changes to per capita tweeting. In these timeseries, an increase from time step $t-1$ to time step $t$ represents not just an increase in overall count of homeless population per capita, but also an increase in the magnitude of population density \textit{growth} in two successive years. Interestingly, cross-correlation coefficients tended to be much lower for most states at the upper end of the distribution when assessing the relationship between changes to per capita homelessness rate and changes to per capita tweeting. Only two states had a cross-correlation greater than 0.75 in magnitude across any potential lag period, Connecticut (0.825 at $t=1$) and Massachusetts (0.796 at $t=1$).  By contrast, geographically large states with relatively low density rankings tended to be closer to zero among the distribution of cross-correlation scores between changes in homelessness rate and changes in tweets, 
including Colorado with density rankings in the range $[25, 31]$; 
Texas, $[26, 31]$; 
Utah, $[40, 42]$; 
Washington, $[11, 14]$; 
Alabama, $[31, 35]$; 
and
Wyoming $[48, 49]$. 
The strength of these two relationships, paired with Massachusetts' outlier status among nationwide homelessness density scores (consistently ranked $1^{\textnormal{st}}$, and Connecticut ranked from $6^\textnormal{th}$ to $8^{\textnormal{th}}$), may indicate that changes to per capita tweet rates are driven by the greater likelihood of an escalation of population growth being more immediately visible to the average Twitter user in smaller, high-density states. 

\begin{table}
    \centering
    \begin{tabular}{|c|c|c|}
    \hline
         State & Coef & $t$  \\
\hline
         \rowcolor{Gray}LA & -0.923 & 1\\
         FL & -0.917 & 1\\
         \rowcolor{Gray}TX & -0.915 & 1 \\
         SC & -0.913 & 4 \\
         \rowcolor{Gray}VA & -0.912 & 1 \\
         NY & 0.906 & 1 \\
        \rowcolor{Gray} OK &  -0.893 & 1 \\
         KY & -0.865 & 1 \\
         \rowcolor{Gray}MA & 0.852 & 1\\
         CO & -0.848 & 1\\
         \hline
         
    \end{tabular}
    \caption{\textbf{Strongest cross-correlation coefficients between per capita homelessness and tweet rates.}}
    \label{tab:crosscorrcoeff}
\end{table}

\begin{figure}
    \centering
    \includegraphics[width=\columnwidth]{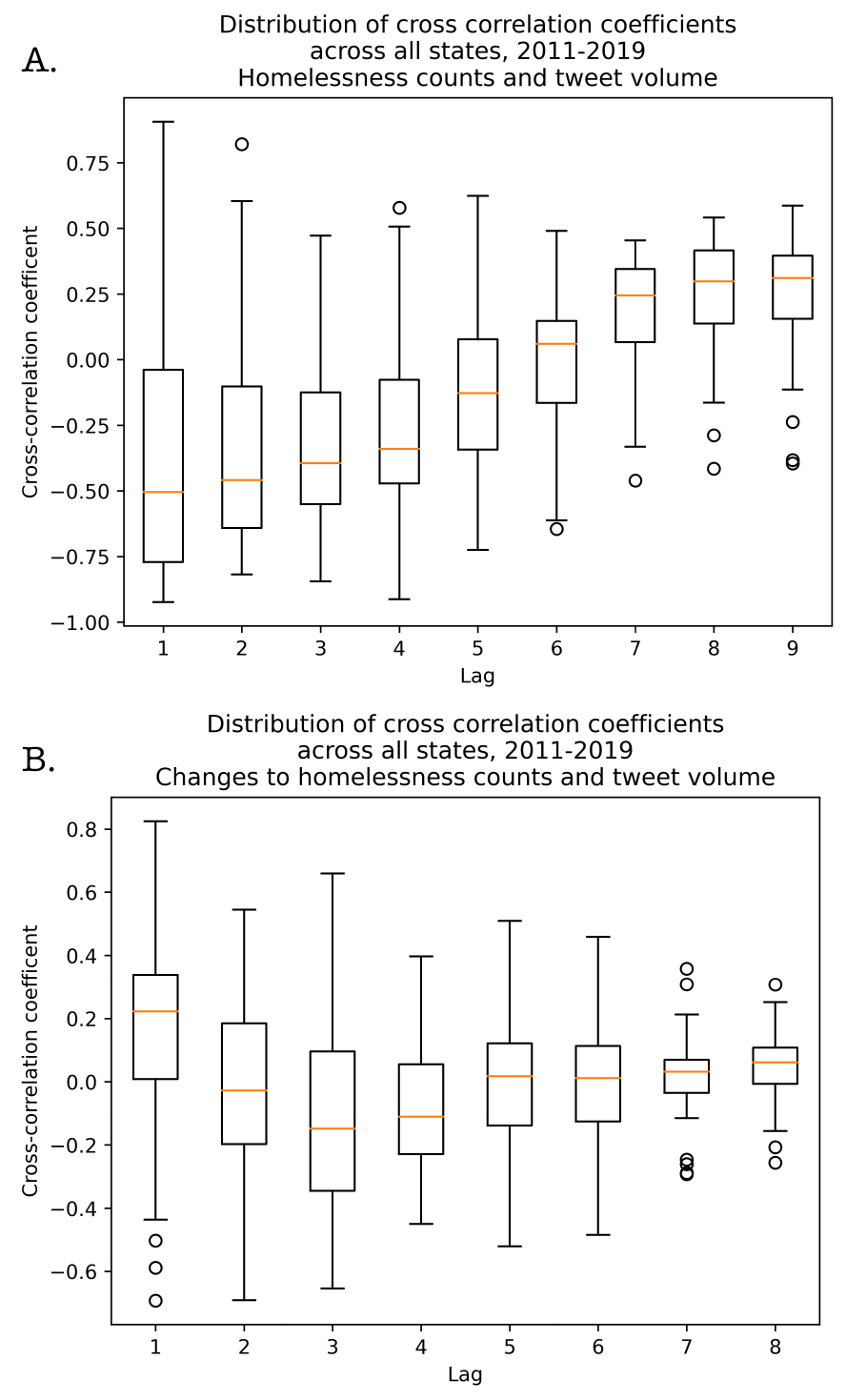}
    \caption{
    \textbf{Distribution of cross-correlation measures between A. per capita total homelessness and tweet volume by state versus B. annualized changes to both.} In the top image, the cross-correlation coefficient of a state approaches 1 at time step $t$ if its per capita tweet count rises and falls in sync with its per capita homeless count after a lag of $t$ years. A cross-correlation coefficient of -1 would indicate perfect anti-correlation: one time series rises as the other falls, and vice versa. In the bottom image, we see the cross-correlation coefficients for changes to per capita tweet  count and homelessness counts, rather than the counts themselves.}
    \label{Fig:crosscorrboxplt}
\end{figure}

\subsection{Homelessness \& Tweet Sentiment}
\label{sec:sentiment}

\subsubsection{Sentiment analysis overview}

Sentiment analysis is a tool widely adopted to gauge the relative positivity or negativity of a text given the sentiment expressed by its constituent pieces, or types (e.g., unique words), weighted by the frequency of each type's appearance (tokens). Though methods to compute a text's composite score vary, one approach has been to rely on sentiment measurements of a set of curated words in frequent use across diverse corpora (i.e., the ``lexicon''), as averaged across scores assigned by trained reviewers who are familiar with the words and their common contexts and uses. This ``bag-of-words'' approach references the score for each unique word in a text that is represented in the lexicon and takes the sum of those scores weighted by the proportion of the text that is represented by each word. Though words may admittedly carry different sentiments to different individuals and across different contexts, several factors contribute to the reliability of this approach for texts, or corpora, of sufficient size: for example, the exclusion of some words from lexicons that are too context-dependent to be reliably scored (e.g., some curse words), the generally-accepted requirement that corpora must contain sufficient words to counterbalance any uncommon use of a small number of types, and the averaging of scores across many reviewers (see Refs.~\cite{hedonometer, sentiment} for more detail).  

For each state, year, month, and state-year, we aggregated all geotagged tweets containing ``homeless'' into a single string, discarded words with a neutral score in the range of $[4.5, 5.5]$~\footnote{
Words with neutral scores that add little to semantic meaning, such as ``of" or ``the", as well as words that are meaningful but have sentiment scores in the middle range, typically comprise the majority of words in a corpus and can drag composite scores toward the middle and mask otherwise-strong sentiment. For this and other reasons, computational linguists tend to exclude them from analysis in order to tune out noise and focus on the meaningful signal generated by the smaller minority of words at either pole of the sentiment scale.}, 
then used the Language assessment by Mechanical Turk (labMT) dictionary to determine compound sentiment scores of large tweet subgroupings (e.g., all tweets geolocated to California in a single month or year), which we then rescaled to a range of possible values of -1 to 1. \cite{hedonometer, sentiment}. 

\subsubsection{Baseline nationwide versus homelessness-related sentiment}

While nearly all corpora exhibited a higher-than-average sentiment score as predicted by the positivity bias observed by  \textcite{polyannaprinciple}, the overall sentiment of tweets containing ``homeless'' was nevertheless lower than the sentiment of overall English-speaking Twitter during the same period and also lower than the sentiment of all US Twitter as documented at \url{hedonometer.org} and \url{https://hedonometer.org/maps.html}, which almost never fell below 5.9 during the period 2010-2019, rescaled to 0.225 under our calculations (see Fig.~\ref{Fig:hedonometer}). For the three years during which the Hedonometer provides an average sentiment score for all US-geotagged tweets (2011, 2012, and 2013), sentiment of US-geotagged tweets containing ``homeless'' falls well below the reference (
See Tab.~\ref{tab:us_sentiment})~\footnote{
It is noteworthy that sentiment of homelessness tweets tends to peak at the end of each year before plummeting at the beginning of the next consecutive year. Although the intervals of available homelessness count data does not permit comparison at the monthly scale, future research could use word shift diagrams to determine whether these cyclical peaks are a result of holiday appeals by homeless-serving agencies, followed by concern expressed for the homeless during cold-weather months in northern states.
}.

\begin{table}[]
    \centering
    \begin{tabular}{|c|c|c|c|}
    \hline
        Year & Average US & Rescaled & Sample Average   \\
         & Sentiment & & Sentiment \\
         \hline
        \rowcolor{Gray}2011 & 5.98 & 0.245 & 0.032\\
        2012 & 5.92 & 0.223 & 0.039\\
        \rowcolor{Gray}2013 & 5.88 & 0.220 &  0.053\\
        \hline
    \end{tabular}
    \caption{\textbf{Comparison of average sentiment of all US tweets versus tweets containing ``homeless''.} Average sentiment of a random sample of 10\% of all English-language tweets from 2011-2013 was filtered for US-geotagged tweets only and published at \url{hedonometer.org}, then rescaled to a range of -1 to 1 for comparison to the average sentiment scores for tweets in the sample of all US-geotagged tweets containing ``homeless''.}
    \label{tab:us_sentiment}
\end{table}

\begin{figure}
    \centering
    \includegraphics[width=\columnwidth]{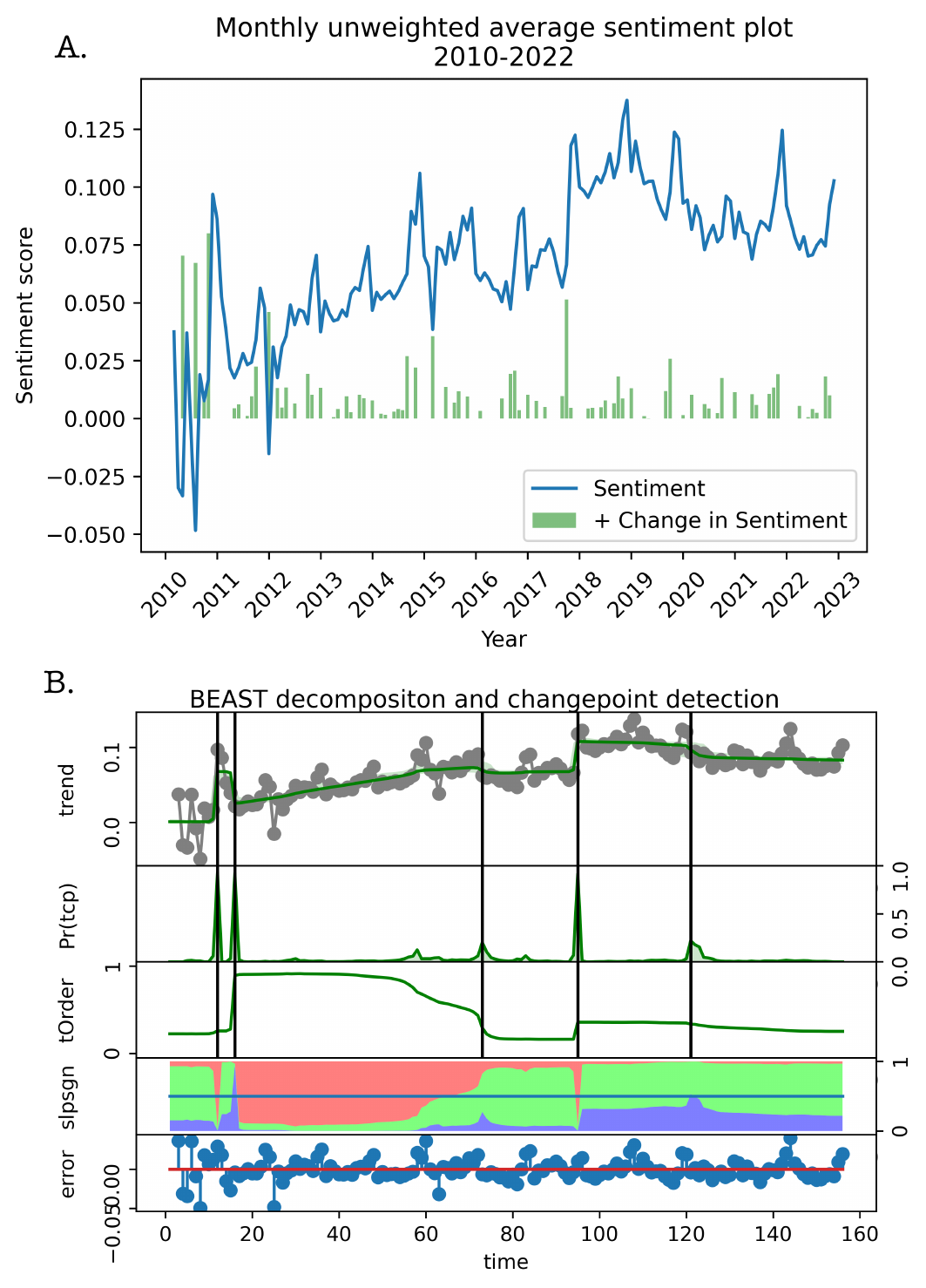}
    \caption{\textbf{B. Bayesian trend change point analysis of A. monthly compound sentiment of all US-originating tweets containing ``homeless'', 2010---2022, overlaid with histogram of magnitude of positive change in successive year, where applicable.} Vertical lines in the lower graphic identify five potential changepoints in sentiment trends detected by the Rbeast package. Note that fluctuations in the early years of the data may reflect greater measurement uncertainty, as total tweet counts prior to 2012 were extremely low. See also Fig. \ref{fig:all_eng_sentiment} at Appendix \ref{sec:measuring-homelessness.storywrangler}.}
    \label{fig:beast}
\end{figure}

\subsubsection{Changepoint detection for national sentiment}
The sentiment of US tweets containing ``homeless'' increased steadily throughout the years 2010--2019 at the same time that odds of usage for our target $n$-gram within US-geotagged Twitter increased. After 2017, the increase in odds and sentiment scores accelerated as the US PiT count documented a change from a nearly decade-long downward trend in overall homeless counts to what has since become 6 years of increasing homelessness. (see Fig. \ref{fig:beast} and \ref{fig:odds}.) Between May and June 2017, all fifty states had published the results of their 2016 Point in Time Counts, and in December 2017, HUD released Part I of its Annual Homelessness Assessment report (AHAR) to Congress, where it reported the nationwide increase for the first time \cite{ahar2017}. A Python-compatible Bayesian estimator of abrupt change and seasonality, Rbeast's trend change point detection function estimates a 99.7\% probability of a change in sentiment trend at time step 95, or December 2017, the end of three successive months of increase in compound sentiment score \cite{beast}. 

Indeed, the Spearman $\rho$ correlation coefficient between ranked raw national homelessness counts and ranked nationwide sentiment scores is statistically significant ($\rho = -0.68$ at $p = 0.02$). However, this correlation between sentiment and homelessness counts is significant only at the national scale; there is no year in which the correlation coefficient between ranked per capita state-level homelessness counts and ranked state-level compound sentiment scores is statistically significant. Variance across the compound sentiment scores of all 50 states, in fact, was extremely low---less than 0.001 for every year since 2011, and only 0.007 in 2010.

\begin{figure*}
    \centering
    \includegraphics[width=\textwidth]{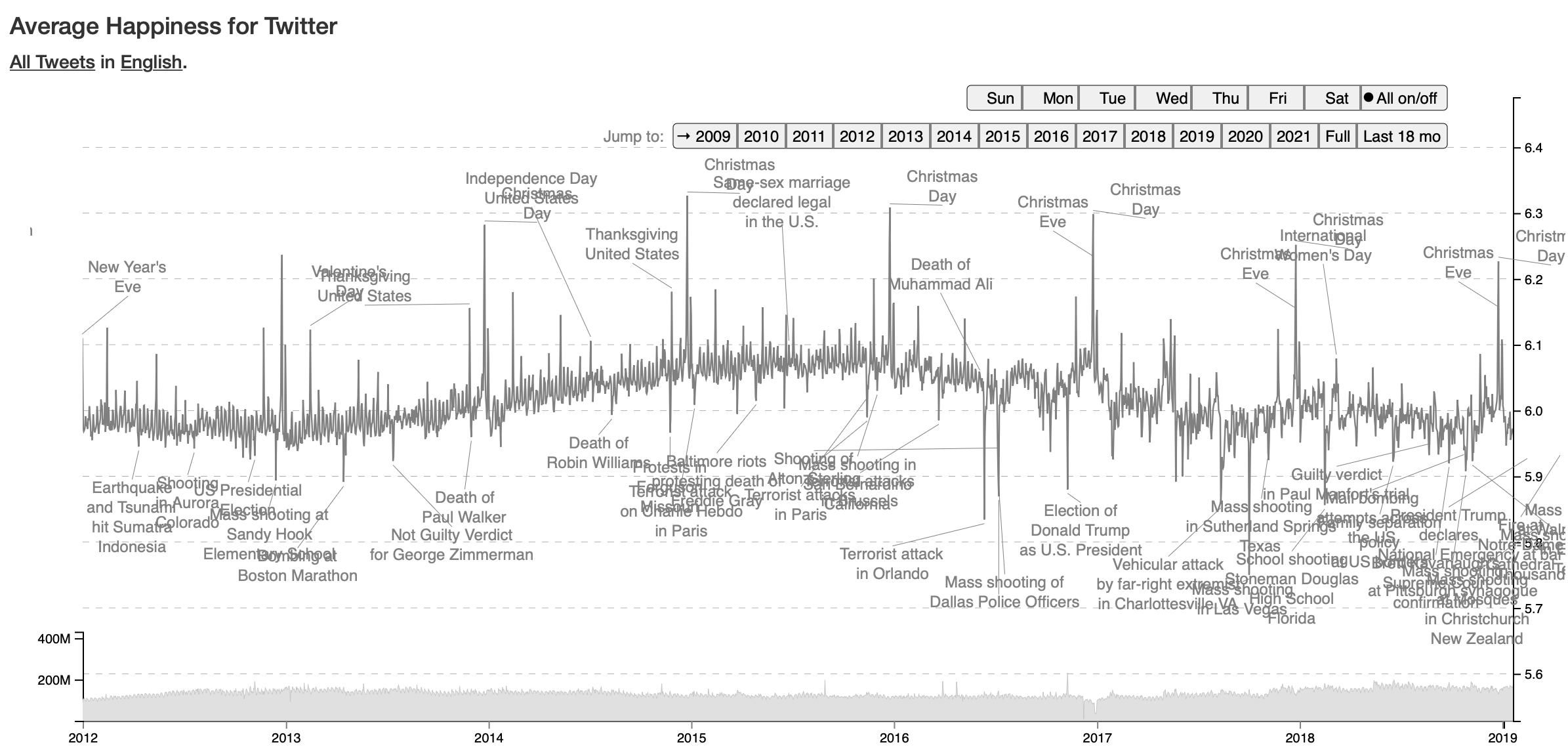}
    \caption{\textbf{Compound sentiment scores across English-language tweets, 2010-2019.} Captured at \url{hedonometer.org}, this time series reflects sentiment scores at the daily scale for a random sample of 10\% of all English-language tweets. Annotations provide context for daily spikes, which typically correspond to annual holidays but can also reflect significant historical events. }
    \label{Fig:hedonometer}
\end{figure*}

The increase in nationwide sentiment score is driven both by a decrease in the number of states with negative scores and also by an increase in magnitude of positive scores among positive-sentiment states. In 2010, roughly half (23, or 46\%) of the states tweeting about homelessness scored below zero on average, with a range of $[-0.4, 0]$, and no states' scores ranged higher than 0.09. After 2013, no state's mean sentiment score is below zero with only two state exceptions in 2015~\footnote{
Colorado and New Mexico, negligibly negative scores of -0.01 and -0.003, respectively}.
By 2019, roughly 2/3 of all states' annualized tweet compound sentiment score was greater than 0.1.

\begin{figure}
    \centering
    \includegraphics[width=\columnwidth]{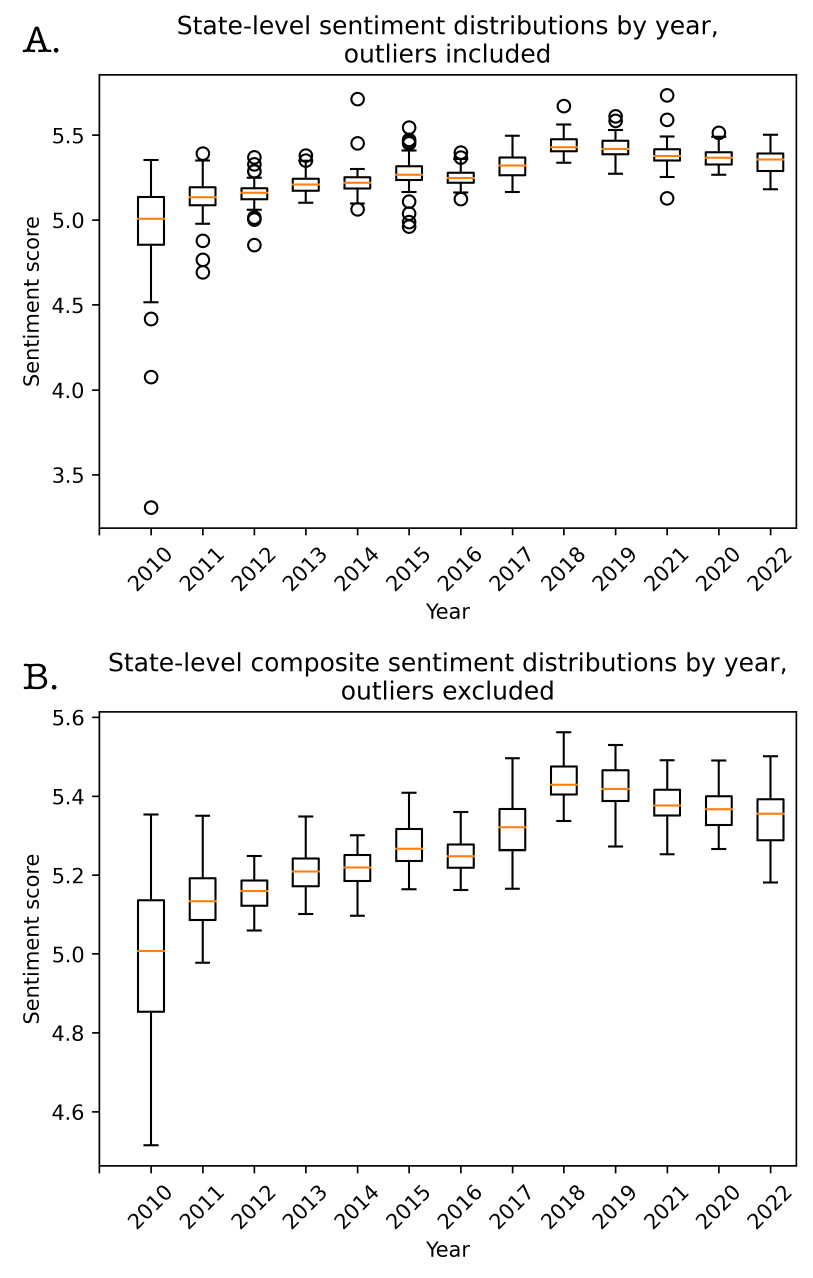}
    \caption{\textbf{Distribution of state compound sentiment scores by year, 2010-2022, A. with and B. without outliers.} Each box-and-whiskers plot visualizes the distribution of all compound sentiment scores for the fifty US states within a single year.}
    \label{fig:sentiment_by_year}
\end{figure}

Curious about the mechanism driving the overall positive trend in the sentiment expressed by tweets containing ``homeless''  even as homelessness increased nationwide, we speculated that either (a) more frequent contact with homelessness could be eliciting among average Twitter users a sense of sympathy, rather than stigma, toward the homeless or (b) any signal of negative sentiment associated with increased homelessness generated by individual Twitter users could be being drowned out by an increase in, for example, appeals and marketing communication by homeless-serving organizations. To investigate further, we began an analysis of the relationship of homelessness rates to tweet content and user type. If, as we hypothesized, homelessness advocacy, service, or policy bodies were responsible for content sentiment change over time, perhaps the proportion of sector-specific language or user accounts could further refine proxy estimates of homelessness rates across the 50 states.

\subsection{Homelessness \& tweet content}

\subsubsection{Corpus curation}
To determine whether rates of homelessness may be distinguishable according to the content of tweets originating from each state, we compared language used in tweets originating from states consistently reporting among the lowest rates of homelessness versus tweets from states with the highest rates of homelessness. This initial analysis narrowly defined high-homelessness states as the six jurisdictions within the top ten rates across all years (California, Washington, Oregon, Nevada, New York, and Hawaii). High-density states included California, Hawaii, and New York as well, in addition to Connecticut, Massachusetts, Maryland, New Jersey, and Rhode Island. At the bottom of each scale, only Kansas and Mississippi ranked in the bottom ten per capita homelessness rates across all years of available data, as contrasted with Alaska, Idaho, Kansas, Montana, North Dakota, Nebraska, New Mexico, South Dakota, Utah, and Wyoming, ten states that remained at the bottom of density rankings consistently. We aggregated all tweets across all years from each cluster of states (i.e., high-homelessness versus low-homelessness, high density versus low density), then ranked by magnitude of difference between the two corpora the relative frequency of each word. The top 50 words occurring more frequently in either corpus relative to the other can be found at Appendix B.

\subsubsection{Sector-specificity and high-homelessness states}

Perhaps not surprisingly, high-homelessness states, when tweeting content containing the word ``homeless'', had higher proportions of words common to housing and homelessness policy and services. As seen in Appendix B, ten industry-specific terms were more common to high-homelessness states, such as ``housing,'' ``employed,'' ``preventing,'' ``encampment,'' and ``youth.'' In contrast, words more commonly associated with low-homelessness states were more likely to be generic words (e.g., ``much,'' ``think,'' ``possible,'' ``want'') or words likely to be associated with crowdsourced and peer-to-peer fundraising (``donate,'' ``paypal,'' ``venmo''), rather than policy solutions.

\label{sec:allotax}
\subsubsection{Allotaxonometry overview}

Allotaxonometry is the comparison of any two complex systems with internally diverse structures~\cite{allotaxonometry}.
For any two text corpora, allotaxonometry with rank-turbulence divergence uses the relative frequency of a word in each respective corpus in order to identify which words contribute the most to the difference between the corpora.
The large, diamond-shaped diagrams in Fig.~\ref{Fig: natlallotax} and in Appendix \ref{sec:measuring-homelessness.allotax}, for example, are each interpreted as a double-histogram in which a word's marker is located along the central dividing line if its rank is equal in both corpora, with the highest-ranked word for both corpora being represented as the apex of the diamond.  If a word appears more often in the reference (or left-hand) corpus, it will appear in the left-hand side of the histogram, as its rank will be lower in that corpus than in the right-hand/comparison corpus. The farther a word appears from the central dividing line, the more extreme the difference in its rank between the two corpora. Words that appear in one corpus but not the other are represented in isolated lines along the lower edges of either side \cite{allotaxonometry, allotaxcode}. 
To the right of the allotaxonometric histogram is a word shift diagram, which ranks the words that contribute the most to the difference between two corpora in reverse order of importance of contribution, from greatest to least importance \cite{wordshifts}. 

\subsubsection{Sector-specificity and nationwide trends}

\begin{figure*}
\centering
    \includegraphics[width=\textwidth]{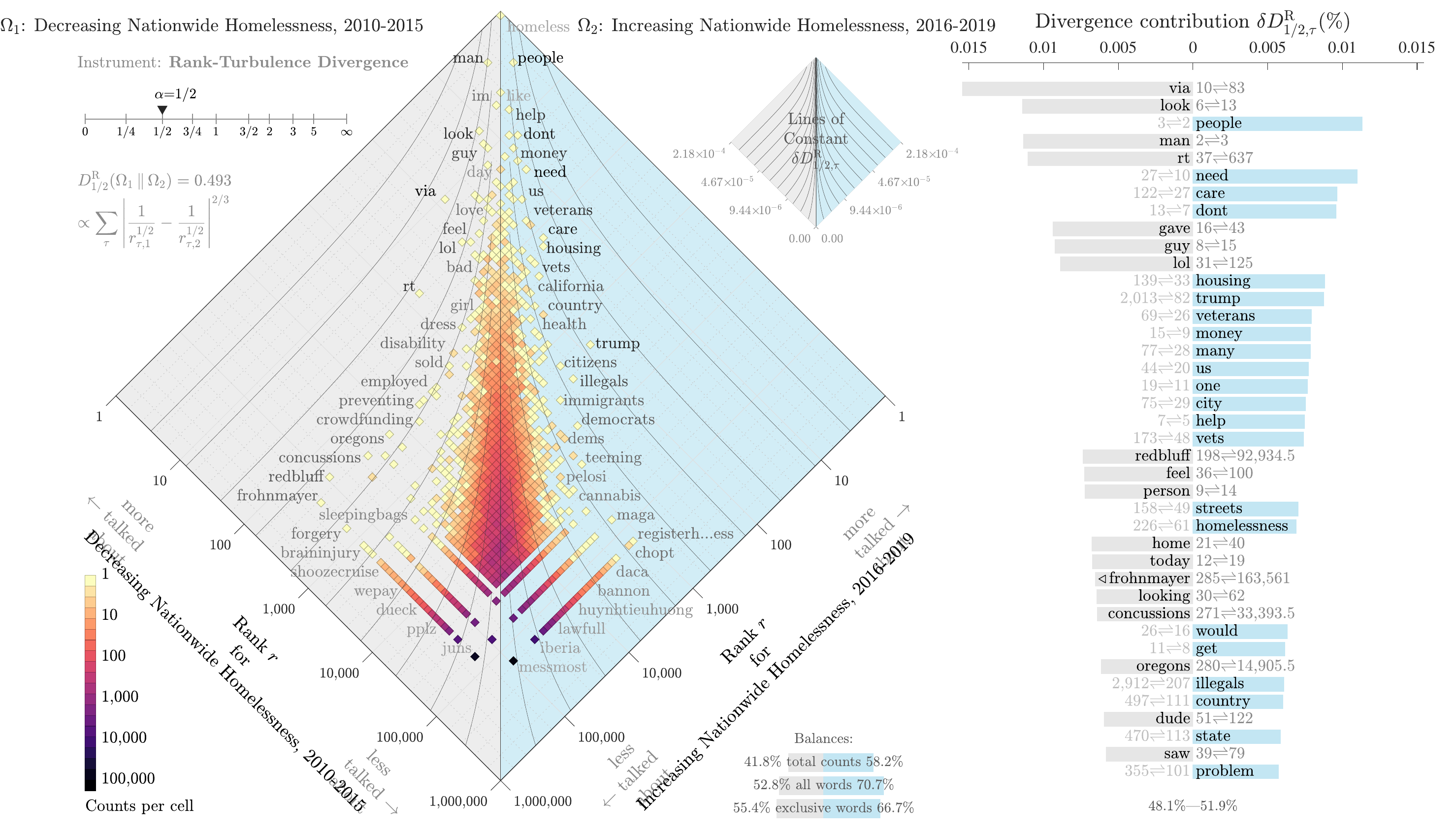}
    \caption{\textbf{Comparisons of tweets from high-homelessness versus low-homelessness
states.} Allotaxonometry is a method that compares two text corpora and, using the relative frequency of a word in each respective corpus, identifies which words contribute the most to the divergence between the corpora. Words appearing to the right or left of the center line appear more frequently in the corpus associated with the side in which they appear. The word shift at right displays the words with the greatest impact on the difference between the two corpora.}
    \label{Fig: natlallotax}
\end{figure*}

We next generated an allotaxonometry diagram comparing tweets from 2010--2015 and from 2016--2022~\footnote{
We included 2016 in the latter period because the first PiT count to register a nationwide increase took place in January 2017, meaning that the increase took place retroactively during 2016}. 
(see Fig.~\ref{Fig: natlallotax}) to see if differences could be captured nationwide between the period of decline in homelessness and the period of nationwide increase. Although the word shift was particularly sensitive to a tuneable alpha parameter value, the words important to distinguishing tweets from the period of increasing homelessness from the period of decreasing homelessness were in every case more likely to reflect a politicization of the issue of homelessness. For example, for tweets posted prior to 2016, apolitical words like ``via'', ``look'', ``man'' ``gave'', and ``guy'' predominate, whereas words like ``housing'', ``trump'', ``veterans'', ``city'', ``vets'', ``homelessness'', ``illegals'', ``country'', and ``state'' characterize the tweets posted 2016 forward (with additional political words~\footnote{e.g., ``daca'', ``registerhomeless'', ``pelosi''} featured, though of lesser importance)~\footnote{
The emphasis on immigration-related words further suggests that the issue of homelessness began to be contextualized within broader political pressure to respond to a reported influx of undocumented immigration to the US via the US-Mexico border, covered at length by American news media starting in 2018 at the end of ex-President Trump's term in office.
}.
This suggests that a change to the overall direction of growth or decline in nationwide rates of homelessness may be visible in the semantic content of an aggregate corpus containing homelessness-related US-geotagged tweets. 

\subsubsection{Sector-specificity and changes in trend polarity}
We then tested to see whether changes in homelessness counts would be similarly visible at the state level by identifying states with variable directional trends in their per capita homelessness counts over time and visualizing allotaxonometry histograms to identify words that were important to any differences between positively-trending periods versus negatively-trending periods. We selected four states---California, Massachusetts, Washington, and Hawaii---which experienced a trend reversal with respect to homelessness volume after at least three consecutive years of monotonic growth or decline in order to ensure a sufficiently robust sample size of tweets within each state-year for comparison. Two states featured a polarity shift from decreasing to increasing trends in annual homelessness rates (California and Washington); the others featured the reverse polarity shift, from increasing to decreasing homelessness (Massachusetts and Hawaii). For California, we aggregated tweets from 2011--2014 (decreasing) to compare against tweets from 2015--2019 and 2022 (increasing); Washington, 2010-2013 (decreasing) versus 2014--2016, 2018 (increasing); Massachusetts, 2012-2014 (increasing) versus 2015--2017 (decreasing); and Hawaii, 2010, 2014-2016 (increasing) versus 2017--2019, 2022 (decreasing). 

In California and Hawaii, sector- and policy-specific language was more prevalent in the period of increasing homelessness. For example, among the California 2015-2019 data, words like ``housing'', ``homelessness'', ``trump'', and ``crisis'' contribute the most to distinguishing the tweets from the California 2010--2014 set, and the word ``democrats'' appears only in the later years' dataset. Similarly, in Hawaii, words like ``appropriation'', ``bill'', ``rights'', ``shelter'', and ``housing'' all appear on the left-hand side of the diagram, which corresponds in Hawaii's case to the period of increasing homelessness rates, and similar language is absent from the right-hand side of the word shift, which depicts important words in distinguishing the period of decreasing homelessness. 

For Massachusetts and Washington, trends observed at the national scale were less clearly reproduced at the state level. While policy and service-sector words were equally balanced in the importance of their contribution to the semantic differences between the two periods in Massachusetts, charity references predominate the period of decreasing homelessness. For example, in the years of increasing Massachusetts homelessness, early-year 1-grams like ``veterans'', ``support'', and ``military'' are counterbalanced by words like ``unemployed'', ``homelessness'', and ``youth'' in the latter years. However, the Massachusetts 2015--2017 tweets are characterized by a prevalence of references to private fundraising events such as ``shoozecruise'', ``charity'', and ``harbordonate'', as well as words potentially related to homeless pets (``freekibble'', ``dogs''). In Washington, by contrast, language typically associated with policy, services, and charity was scarce across both periods, and so the time frames of increasing and decreasing homelessness were not easily distinguishable.  

It is worth noting that trends observed at the national scale are not present at the state level at the unit of individual years (or ``state-years''). To test whether language would differ significantly between increasing and decreasing state-years, we aggregated into two separate corpora (1) all tweets from states in years $x$ where the rate had increased from the year $x-1$ and (2) all tweets from states in which the rate had decreased. We then generated an allotaxonometry diagram comparing the corpora. Approximately equal proportions of sector-specific terms characterize both corpora, and neither corpus was distinguished by references to charitable efforts or direct appeals. It is unclear whether the corpora's similarity is a result of distributions of state-years from the period of nationwide increasing homelessness across both corpora, indicating that national discourse during the post-2017 era may in some cases overcome semantic differences that would otherwise be present, or because of the presence of anomalous single-year polarity changes within otherwise-monotonic, reverse-polar trends, indicating that changes must persist several years in order to effect semantic changes in social media language.  

\subsection{User Type, Tweet Volume \& Content}

Multiple findings of the analysis above recommend a closer inspection of the relationship between user type and tweet volume, content, and sentiment with respect to homelessness. The observation by prior literature of content and value differences between homeless and non-homeless social media users suggests that differences observed in sentiment and content may be the result of increased activity by altruistic Twitter users and/or advocate networks. Thus, we finally reviewed homelessness-related Twitter use patterns by account type as a potential means of signal detection for real-world changes to local homelessness volume.

While a full examination of the impact of account type is outside the scope of our immediate investigations, we nevertheless undertook a preliminary analysis of frequency of posts on homelessness by user account to explore, in a general way, whether unusually high-frequency posters in high-homelessness versus low-homelessness states were individuals or entities, and if the latter, what type of entities. We began with a sample of tweets from the 100 state-years with the highest reported homelessness density ($N_{h}=313,311$ from 134,616 accounts) versus the 100 state-years with the lowest reported homelessness density ($N_{l}=19,966$ from 11,078 accounts). Because account type is not among the metadata provided by Twitter's API, we manually labeled the accounts with the highest number of tweets in each corpus as Individual versus Entity accounts, indicating a sub-type where relevant, e.g., Agency, News, or Legal for Entities and Politician or Journalist for Individuals. 

\begin{figure*}
    \includegraphics[width= 6 in]{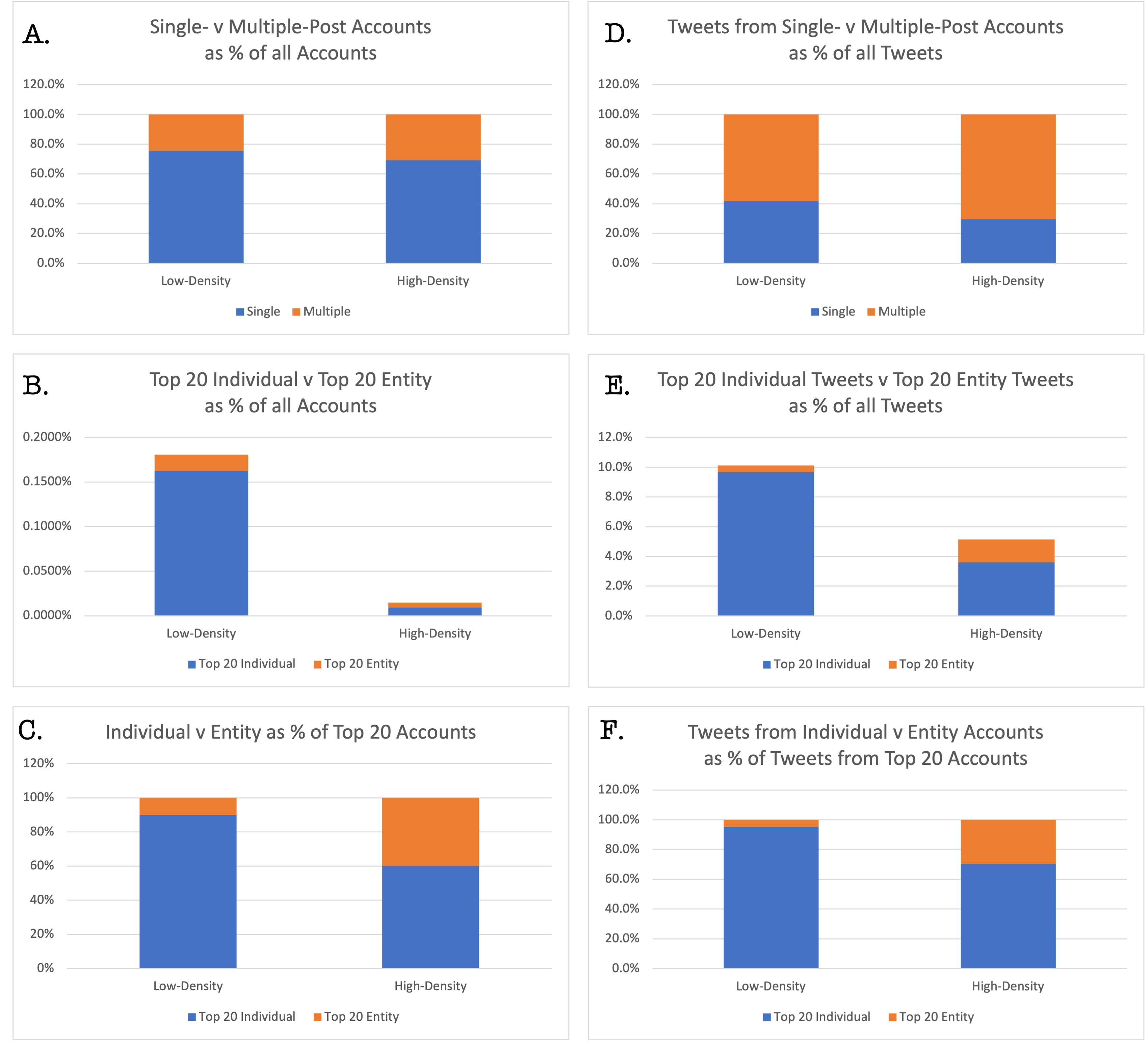}
    \caption{\textbf{Comparison of the ratios of user account types in low-density versus high-density state-years.} C. Entity accounts comprise a greater percentage of the top 20 accounts posting most frequently about homelessness in high-density versus low-density states, and A. accounts tend to tweet more than once about homelessness in high-density states.}
\end{figure*}

\subsubsection{Low- versus high-density state-years}

Our first finding was that the relative prevalence of users who posted only once about homelessness (``single-post'' accounts) versus users who posted more than once within the dataset (``multi-post'' accounts) differed significantly by homelessness density (75.5\% versus 24.5\% in low-density state-years, 69.2\% versus 30.8\% in high-density state-years), indicating that accounts were somewhat more likely to post more than once about homelessness in high-density states. Additionally, tweets from those multi-post accounts comprised 70.3\% of all tweets from high homelessness-density states, as contrasted with only 58.1\% of all tweets from low-density states. 

The difference in sample sizes from high-density and low-density states resulted in top 20 accounts generating a much smaller percentage of overall tweets in low-density states (0.18\% versus 0.01\%). The ratio of the percentage represented by top 20 individual accounts to top 20 entity accounts, however, is 9:1 in low-density states, or six times that in high-density states, 3:2.

Individual accounts represented a significantly greater proportion of top 20 accounts in low-density states (90\%) versus high-density states (60\%). The percentage of top 20 tweets represented by those individual accounts also differed significantly by density (95.4\% in low-density state-years versus 70.2\%). 

In sum, in higher-density state-years, a greater proportion of the users tweeting about homelessness are doing so more than once and generating a higher percentage of the total tweet corpus. High-frequency posters are far more likely to be individuals than entities in low-density state-years, and the proportion of tweets they generate within the top 20 user tweet corpus is also much higher in low-density state-years, suggesting a potential increase in entity-account online presence as states increase in homelessness density. 

\begin{table}[]
    \centering
    \begin{tabular}{|c|c|c|}
\hline
User rank & Tweet Count & Account type \\
\rowcolor{Gray} 1  &  5865  &  Individual \\
2  &  1653  &  Entity \\
\rowcolor{Gray} 3  &  1397  &  Individual \\
4  &  784  &  News \\
\rowcolor{Gray} 5  &  487  &  Individual \\
6  &  483  &  News \\
\rowcolor{Gray} 7  &  480  &  Individual \\
8  &  480  &  News \\
\rowcolor{Gray} 9  &  427  &  Individual \\
10  &  425  &  Individual \\
\rowcolor{Gray} 11  &  402  &  Individual \\
12  &  401  &  Agency \\
\rowcolor{Gray} 13  &  397  &  Individual \\
14  &  394  &  Individual \\
\rowcolor{Gray} 15  &  383  &  Individual \\
16  &  352  &  Legal \\
\rowcolor{Gray} 17  &  347  &  Legal \\
18  &  334  &  Individual \\
\rowcolor{Gray} 19  &  323  &  Individual \\
20  &  313  &  News \\
\hline
\end{tabular}
\caption{\textbf{Tweet counts (for tweets containing ``homeless'') and account type classification for high-frequency users in high-density state-years.}}
\end{table}

\begin{table}[]
    \centering
    \begin{tabular}{|c|c|c|}
\hline
User rank & Tweet Count & Account type \\
\rowcolor{Gray} 1  &  523  &  Individual \\
2  &  442  &  Individual \\
\rowcolor{Gray} 3  &  150  &  Individual \\
4  &  84  &  Individual \\
\rowcolor{Gray} 5  &  81  &  Individual \\
6  &  74  &  Individual \\
\rowcolor{Gray} 7  &  71  &  Individual \\
8  &  68  &  Individual \\
\rowcolor{Gray} 9  &  66  &  Individual \\
10  &  65  &  Individual \\
\rowcolor{Gray} 11  &  57  &  Legal \\
12  &  49  &  Individual \\
\rowcolor{Gray} 13  &  48  &  Individual \\
14  &  46  &  Individual \\
\rowcolor{Gray} 15  &  36  &  News \\
16  &  36  &  Individual \\
\rowcolor{Gray} 17  &  33  &  Individual \\
18  &  32  &  Individual \\
\rowcolor{Gray} 19  &  32  &  Individual \\
20  &  29  &  Individual \\
\hline
\end{tabular}
\caption{\textbf{Tweet counts (for tweets containing ``homeless'') and account type classification for high-frequency users in low-density state-years.}}
\end{table}

\subsubsection{Negatively- versus positively-sentimented state-years}

We then reproduced this approach with a sample of tweets from the 30 negatively-sentimented state-years ($N_{n} = 5,593$ generated by 2574 users) and a sample of tweets from the 30 most positively-sentimented state-years ($N_{p} = 3,5820$ generated by 17482 users) for comparison. 

\begin{figure*}
    \includegraphics[width=6 in]{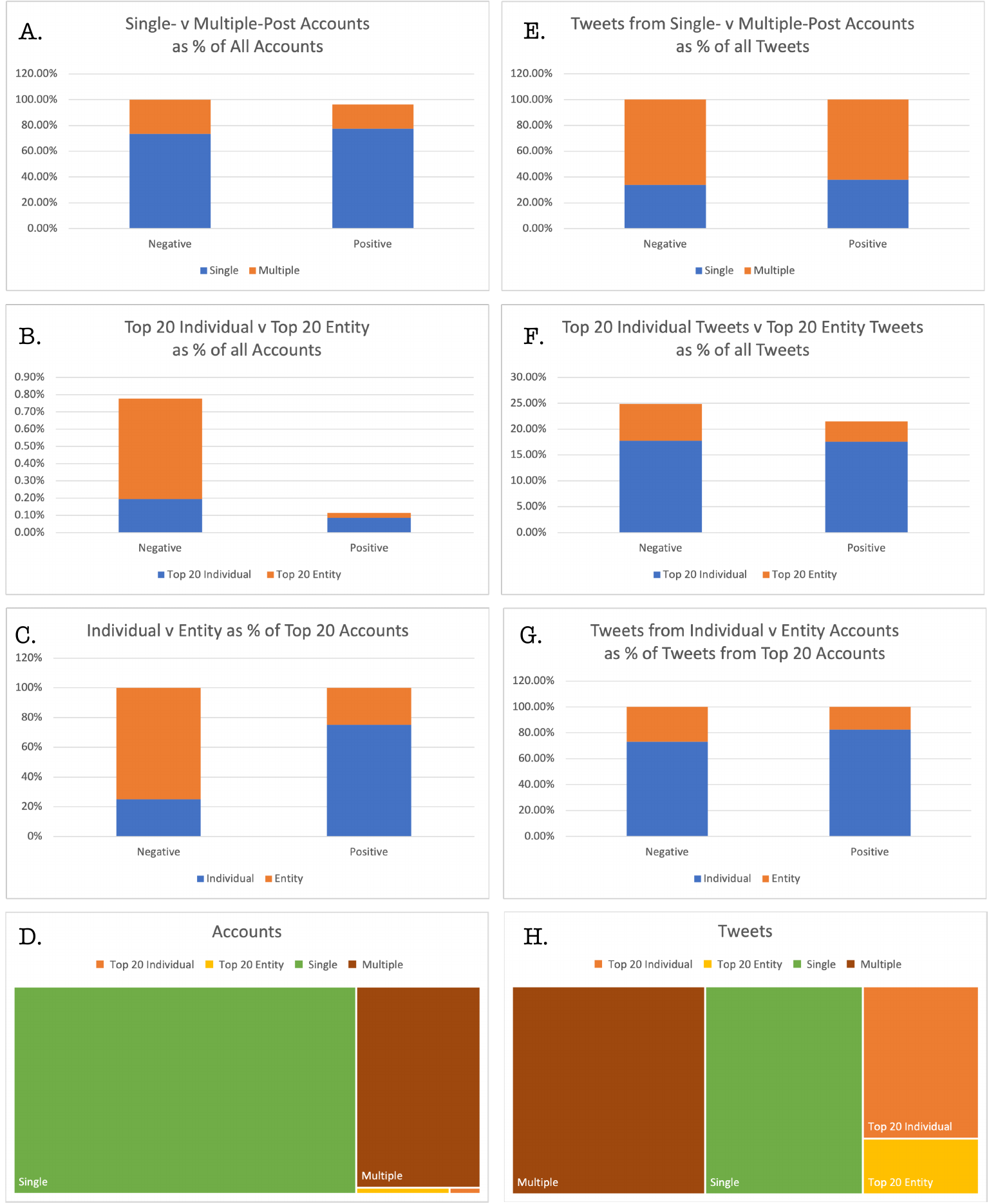}
    \caption{\textbf{Comparison of the ratios of user account types in positively- versus negatively-sentimented state-years.} C. Entity accounts comprise a much smaller percentage of top 20 accounts posting most frequently about homelessness in positively-sentimented states. F. Tweets from entity accounts, regardless of frequency of posting, constitute a higher percentage of tweets in negatively-sentimented states as well.}
\end{figure*}

Overall, single-tweet accounts comprise a much larger percentage of the accounts tweeting about homelessness across all state-years (positive and negative), but the number of tweets generated by multiple-tweet accounts represents a larger percentage of all tweets~\footnote{
Note that Top 20 Individual and Top 20 Entity accounts are all multiple-post accounts}.
We found that the relative contribution of single-post accounts again differed significantly by corpus sentiment, though less so than by density. Single-post accounts represented 73.4\% and 77.6\% of all accounts in positively-sentimented versus negatively-sentimented state-years, respectively; multi-post accounts, 26.6\% and 22.4\%. Similarly, the proportion of tweets generated by a single-post account to those generated by multi-post accounts differed significantly by sentiment, or 33.8\% versus 37.9\% and 66.2\% versus 62.1\%.

Despite the fact that the number of accounts represented in negatively-sentimented states is only 14.7\% of accounts in positively-sentimented states, top 20 accounts comprise an 24.8\% of all accounts in the former and 21.3\% in the latter. The ratio of top 20 entities to individuals as expressed as percentage of all accounts is twice as large in negatively-sentimented (3:1, or 0.58\% versus 0.19\%) as it is in positively-sentimented states (1:3, or 0.03\% and 0.09\%). Thus, user accounts tweeting about homelessness in negatively-sentimented states are more likely to be individuals than entities, while the opposite is true in positively-sentimented states. Tweets from those top 20 entity accounts comprise over 1.7 times the percentage of all tweets in negatively-sentimented states (6.7\%) as they do in positively-sentimented states (3.8\%), versus tweets from top 20 individual accounts (18.2\% in negatively-sentimented states, 17.6\% otherwise). 

\begin{table}
    \centering
    \begin{tabular}{|c|c|c|}
\hline
     User rank & Tweet Count & Account type\\
     \hline
\rowcolor{Gray}1  &  1857  &  Individual\\
2  &  1631  &  Individual\\
\rowcolor{Gray}3  &  867  &  Individual\\
4  &  394  &  Individual\\
\rowcolor{Gray}5  &  374  &  Entity\\
6  &  332  &  Individual\\
\rowcolor{Gray}7  &  327  &  Agency\\
8  &  273  &  Entity\\
\rowcolor{Gray}9  &  261  &  Individual\\
10  &  231  &  Individual\\
\rowcolor{Gray}11  &  221  &  Agency\\
12  &  195  &  Individual\\
\rowcolor{Gray}13  &  168  &  Entity\\
14  &  150  &  Individual\\
\rowcolor{Gray}15  &  120  &  Individual\\
16  &  117  &  Individual\\
\rowcolor{Gray}17  &  70  &  Individual\\
18  &  69  &  Individual\\
\rowcolor{Gray}19  &  63  &  Individual\\
20  &  62  &  Individual\\
\hline
\end{tabular}
\caption{\textbf{Tweet counts (for tweets containing ``homeless'') and account type classification for high-frequency users in positively-sentimented state-years}}
\end{table}

\begin{table}
    \centering
    \begin{tabular}{|c|c|c|}
\hline
User rank & Tweet Count & Account type \\
\hline
\rowcolor{Gray} 1  &  796  &  Individual \\
2  &  131  &  Individual \\
\rowcolor{Gray} 3  &  46  &  Individual \\
4  &  45  &  Agency \\
\rowcolor{Gray} 5  &  44  &  News \\
6  &  39  &  News \\
\rowcolor{Gray} 7  &  34  &  News \\
8  &  25  &  Legal \\
\rowcolor{Gray} 9  &  24  &  Individual \\
10  &  23  &  News \\
\rowcolor{Gray} 11  &  20  &  Individual \\
12  &  20  &  News \\
\rowcolor{Gray} 13  &  19  &  News \\
14  &  18  &  News \\
\rowcolor{Gray} 15  &  18  &  News \\
16  &  18  &  News \\
\rowcolor{Gray} 17  &  18  &  News \\
18  &  17  &  News \\
\rowcolor{Gray} 19  &  17  &  News \\
20  &  17  &  News \\
\hline
\end{tabular}
\caption{\textbf{Tweet counts (for tweets containing ``homeless'') and account type classification for high-frequency users in negatively-sentimented state-years}}
\end{table}

Entity accounts dominate individual accounts among the top 20 highest-frequency users in negatively-sentimented states (75\% versus 25\%); the opposite is true in positively-sentimented states (25\% versus 75\%). Additionally, although the proportion of all tweets represented by individual top 20 high-frequency accounts among all tweets differs by only 1.4\% in negatively- versus positively-sentimented state-years (18.2\% and 17.6\%), the proportion of tweets generated by individuals versus entities among the top 20 accounts' tweets is lower in negatively-sentimented states (73.2\% versus 82.4\%). Thus, while there are more single-post accounts than multi-post accounts across both positively- and negatively-sentimented states and more individuals than entities among the top 20 in positively-sentimented states, a handful of high-frequency accounts generate a greater percentage of tweets in the negatively-sentimented corpus. The presence of a higher proportion of individuals among positively-sentimented high-frequency accounts, and a higher proportion of top 20 tweets being generated by those individuals, does not increase their contribution as much to tweets overall as might be expected, while a decrease in entity presence among the positive state-years yields both a lower proportion of top 20 tweets and a decreased presence among overall tweets in the corpus.  

\section{Conclusions}

Within a given year, several social media indicators may generate a signal that correlates to changes in a jurisdiction's rate or volume of homelessness, including tweet rates, semantic content, user behavior, and ratios of different account types tweeting about homelessness. Some statistically significant nonparametric rank-rank correlations exist between tweet volume ranked by state within a given year, for example, and those state's comparative ranked densities of homelessness. Additionally, the likelihood of tweets to contain the word ``homeless'', as well as the average sentiment of those tweets, is sensitive to, and revealing of, national trends. However, the variability of population density distributions between and within states translates into different probabilities of real-world experiences of, and encounters with, homelessness by the typical social media user within a given jurisdiction. For example, correlations exist across years within certain---but not all---states between annualized tweet and homelessness values, particularly exceptionally high-density, high-homelessness states with high variance among annual state-level estimates. 
More research is needed to better understand why tweet rates generated by other high-density, high-variability states do not correlate with changes to homelessness estimates by year, and why the strongest cross-correlation coefficients of tweet volume to homelessness rates are negative for many southern states, yet positive in two New England states. 

Moreover, preliminary findings suggest that some retrospective or real-time trend detection may be possible through text analysis. 
Further research, which would require tools for geolocation and/or user-type classification, is needed to determine whether findings are generalizable at scale. 
A deeper exploration of the mechanisms underlying a typical social media user's understanding about conditions within their community---not only real-world interactions with people experiencing homelessness, but also indirect sources of information, such as news and social media---is critical to modeling and predicting the relationship of some more complex public health phenomena to online behavior, particularly in cases, as with homelessness, real-world experiences are heterogeneous and context-sensitive.  

In sum, a measurement consistent with established measures like the PiT cannot be straightforwardly obtained from Twitter volume, content, or sentiment, though comparisons between and among jurisdictions may be reasonably within grasp. 
Through a combination of real-time analytic techniques, including computational linguistics, social media data may yet provide insight into the mechanisms translating local homelessness measurements into online behavior. The cost of Twitter API access, however, which has increased dramatically since this project began, makes real-time data acquisition cost-prohibitive for many interested stakeholders with limited resources, such as smaller advocacy organizations or watchgroups. Extension of this research to other, less costly social media platforms is recommended (1) to ensure the robustness of these findings across fora and (2) to identify lower-cost real-time datasets that would be useful for analysis. Additionally, research at a more or less granular level---e.g., the city or county level within a single country, or at the nation-level internationally---will be critical to determine whether correlations between homelessness and tweet volume persist in environments characterized by greater heterogeneity and rank-turbulence with respect to homelessness rates and density. However, the challenge of identifying valid, real-time ground-truth datasets at these other levels against which to compare social media observations remains a difficult problem, given methods used to estimate homelessness, at least in the US. 

\subsubsection{Limitations of the study}

Each dataset presented unique challenges to the validity and generalizability of our results. First and most importantly, the Point-in-Time Count relies on non-uniform data collection practices that are of variable quality, such as manual counts of unsheltered homeless who are identified outdoors during a single-night annual count the last week of January. Because these practices cannot be expected to perform comparatively to scale across jurisdictions with varying population densities, available volunteer base, outreach project data quality, and distributions of rural/urban centers, states, HUD permits different approaches to estimating the size of the unsheltered population, including census-based approaches, non-random sampling with extrapolation, or a combination of those approaches \cite{pitmethodguide}. Moreover, even if data collection practices were standardized, the annualized time scale makes it impossible to view population flow dynamics during later months of the year and may underestimate regional needs in cold-weather states if the homeless are migrating to warmer locations seasonally. Consequently, any conclusions drawn regarding the relationship of homelessness rates to Twitter activity, even if validated against these counts, should considered within a broader context alongside other suspected proxies of a state's actual level of homelessness, such as eviction or 
unemployment rates 
or housing affordability~\footnote{
The percent of household income spent, on average, for housing costs}.

Secondly, geotagged English-language data represents only a small fraction of overall Twitter activity generated in the US, which may contain hidden biases based on, for example, distributed proportions of multilingual or limited English proficiency Twitter users.  Different state or regional norms around data and privacy protection with respect to social media may also affect the proportional representation of a jurisdiction within geotagged data. It is also well-documented that Twitter users are not a representative or random sample of United States residents \cite{happiness}. A random sample of all US-originating tweets with and without geotagging, filterable for multiple languages' customary analogues to the 1-gram ``homeless'', would be a preferable dataset, but interpolating valid user location is a difficult problem often requiring time-intensive manual coding \cite{wearevisibleshort}, and monthly caps on the volume of tweet and user account data downloads made this problem beyond the scope of the present research.

Relatedly, homelessness, an often-sensitive political issue, may attract a volume and sentiment of social media communication that corresponds to factors and agendas other than observed homelessness rates. Political noise may mask an otherwise-detectable signal based on average citizens' observations of homelessness in their communities. For this reason, further analysis by user type (e.g., advocacy organization, news media, private individual account) is critical to better understanding the relationship of various actors, their online speech, and aggregate measures of sentiment, volume, and content, and how it is distinguishable from speech signals generated within tweets from the average individual.

Finally, tweets were assumed to originate within the state identified by the geoid dictionary. Artificial bot amplification of tweet volume on homelessness was similarly assumed to be equally likely across all fifty states and was not specifically addressed. The impact of bots on homelessness-related Twitter communication should be an area of future research.

\section{Code and Data}
\label{sec:measuring-homelessness.code}

The code described herein can be found in the public Google Colab folder here: \url{https://drive.google.com/file/d/1va4dWRZ6RPT089rcTkrE5fvK5hT0nksV/view?usp=sharing}.

We accessed the Twitter dataset described above by submitting the following query parameters to Twitter API v. $2.0$: 

query = 'homeless place\_country:US has:geo'
start\_time = '2010-01-01T00:00:00Z'
end\_time = '2022-12-31T23:59:00Z'


\acknowledgments
We are grateful for support from 
the University of  Vermont Complex Systems Center, 
the MassMutual Center of Excellence in Complex Systems and Data Science, 
and discussions with colleagues from the Computational Story Lab.


\bibliography{\filenamebase.bib}

\begin{thebibliography}{38}%
\makeatletter
\providecommand \@ifxundefined [1]{%
 \@ifx{#1\undefined}
}%
\providecommand \@ifnum [1]{%
 \ifnum #1\expandafter \@firstoftwo
 \else \expandafter \@secondoftwo
 \fi
}%
\providecommand \@ifx [1]{%
 \ifx #1\expandafter \@firstoftwo
 \else \expandafter \@secondoftwo
 \fi
}%
\providecommand \natexlab [1]{#1}%
\providecommand \enquote  [1]{``#1''}%
\providecommand \bibnamefont  [1]{#1}%
\providecommand \bibfnamefont [1]{#1}%
\providecommand \citenamefont [1]{#1}%
\providecommand \href@noop [0]{\@secondoftwo}%
\providecommand \href [0]{\begingroup \@sanitize@url \@href}%
\providecommand \@href[1]{\@@startlink{#1}\@@href}%
\providecommand \@@href[1]{\endgroup#1\@@endlink}%
\providecommand \@sanitize@url [0]{\catcode `\\12\catcode `\$12\catcode
  `\&12\catcode `\#12\catcode `\^12\catcode `\_12\catcode `\%12\relax}%
\providecommand \@@startlink[1]{}%
\providecommand \@@endlink[0]{}%
\providecommand \url  [0]{\begingroup\@sanitize@url \@url }%
\providecommand \@url [1]{\endgroup\@href {#1}{\urlprefix }}%
\providecommand \urlprefix  [0]{URL }%
\providecommand \Eprint [0]{\href }%
\providecommand \doibase [0]{https://doi.org/}%
\providecommand \selectlanguage [0]{\@gobble}%
\providecommand \bibinfo  [0]{\@secondoftwo}%
\providecommand \bibfield  [0]{\@secondoftwo}%
\providecommand \translation [1]{[#1]}%
\providecommand \BibitemOpen [0]{}%
\providecommand \bibitemStop [0]{}%
\providecommand \bibitemNoStop [0]{.\EOS\space}%
\providecommand \EOS [0]{\spacefactor3000\relax}%
\providecommand \BibitemShut  [1]{\csname bibitem#1\endcsname}%
\let\auto@bib@innerbib\@empty
\bibitem [{\citenamefont {{United States Department of Housing and Urban
  Development}}(2023)}]{pitestimates}%
  \BibitemOpen
  \bibfield  {author} {\bibinfo {author} {\bibnamefont {{United States
  Department of Housing and Urban Development}}},\ }\href
  {https://www.huduser.gov/portal/sites/default/files/xls/2007-2021-PIT-Counts-by-CoC.xlsx}
  {\bibinfo {title} {2007 - 2022 {P}oint in {T}ime {E}stimates by {C}o{C}
  [{D}ata set]}} (\bibinfo {year} {2023})\BibitemShut {NoStop}%
\bibitem [{\citenamefont {{Health Care for the Homeless Maryland}}()}]{hchmd}%
  \BibitemOpen
  \bibfield  {author} {\bibinfo {author} {\bibnamefont {{Health Care for the
  Homeless Maryland}}},\ }\href
  {http://www.hchmd.org/homelessness-makes-you-sick} {\bibinfo {title}
  {Homelessness {M}akes {Y}ou {S}ick}}\BibitemShut {NoStop}%
\bibitem [{\citenamefont {{United States Interagency Council on
  Homelessness}}(2015)}]{openingdoors}%
  \BibitemOpen
  \bibfield  {author} {\bibinfo {author} {\bibnamefont {{United States
  Interagency Council on Homelessness}}},\ }\href
  {https://www.usich.gov/resources/uploads/asset\_library/USICH\_OpeningDoors\_Amendment2015\_FINAL.pdf}
  {\bibinfo {title} {Opening {D}oors: {F}ederal {S}trategic {P}lan to {P}revent
  and {E}nd {H}omelessness as amended in 2015}} (\bibinfo {year}
  {2015})\BibitemShut {NoStop}%
\bibitem [{\citenamefont {Almquist}\ \emph {et~al.}(2020)\citenamefont
  {Almquist}, \citenamefont {Helwig},\ and\ \citenamefont {You}}]{CoCcounty}%
  \BibitemOpen
  \bibfield  {author} {\bibinfo {author} {\bibfnamefont {Z.}~\bibnamefont
  {Almquist}}, \bibinfo {author} {\bibfnamefont {N.}~\bibnamefont {Helwig}},\
  and\ \bibinfo {author} {\bibfnamefont {Y.}~\bibnamefont {You}},\ }\bibfield
  {title} {\bibinfo {title} {Connecting {C}ontinuum of {C}are point-in-time
  homeless counts to {U}nited {S}tates {C}ensus areal units},\ }\href
  {https://doi.org/https://doi.org/10.1080/08898480.2019.1636574} {\bibfield
  {journal} {\bibinfo  {journal} {Mathematical Population Studies}\ }\textbf
  {\bibinfo {volume} {27}},\ \bibinfo {pages} {46} (\bibinfo {year}
  {2020})}\BibitemShut {NoStop}%
\bibitem [{\citenamefont {Swenson}(2022)}]{wapo}%
  \BibitemOpen
  \bibfield  {author} {\bibinfo {author} {\bibfnamefont {K.}~\bibnamefont
  {Swenson}},\ }\bibfield  {title} {\bibinfo {title} {America’s first
  homelessness problem: Knowing who is actually homeless},\ }\href
  {https://www.washingtonpost.com/dc-md-va/2022/08/24/homeless-seattle-hud-statistics/}
  {\bibfield  {journal} {\bibinfo  {journal} {Washington Post}\ } (\bibinfo
  {year} {2022})}\BibitemShut {NoStop}%
\bibitem [{\citenamefont {Hu}\ \emph {et~al.}(2019)\citenamefont {Hu},
  \citenamefont {Chancellor},\ and\ \citenamefont
  {Choudhury}}]{clusterlitreview}%
  \BibitemOpen
  \bibfield  {author} {\bibinfo {author} {\bibfnamefont {A.}~\bibnamefont
  {Hu}}, \bibinfo {author} {\bibfnamefont {S.}~\bibnamefont {Chancellor}},\
  and\ \bibinfo {author} {\bibfnamefont {M.~D.}\ \bibnamefont {Choudhury}},\
  }\bibfield  {title} {\bibinfo {title} {Characterizing {H}omelessness
  {D}iscourse on {S}ocial {M}edia},\ }\href
  {https://doi.org/https://doi.org/10.1145/3290607.3313057} {\bibfield
  {journal} {\bibinfo  {journal} {Extended Abstracts of the 2019 CHI Conference
  on Human Factors in Computing Systems}\ ,\ \bibinfo {pages} {1}} (\bibinfo
  {year} {2019})}\BibitemShut {NoStop}%
\bibitem [{\citenamefont {Koepfler}\ \emph {et~al.}(2013)\citenamefont
  {Koepfler}, \citenamefont {Shilton},\ and\ \citenamefont
  {Fleischmann}}]{wearevisiblefull}%
  \BibitemOpen
  \bibfield  {author} {\bibinfo {author} {\bibfnamefont {J.}~\bibnamefont
  {Koepfler}}, \bibinfo {author} {\bibfnamefont {K.}~\bibnamefont {Shilton}},\
  and\ \bibinfo {author} {\bibfnamefont {K.}~\bibnamefont {Fleischmann}},\
  }\bibfield  {title} {\bibinfo {title} {A stake in the issue of homelessness:
  Identifying values of interest for design in online communities},\ }\href
  {https://doi.org/https://doi.org/10.1145/2482991.2482994} {\bibfield
  {journal} {\bibinfo  {journal} {Proceedings of the 6th International
  Conference on Communities and Technologies - C\&T ’13}\ ,\ \bibinfo {pages}
  {36}} (\bibinfo {year} {2013})}\BibitemShut {NoStop}%
\bibitem [{Note1()}]{Note1}%
  \BibitemOpen
  \bibinfo {note} {Follow-up work moved away from account labels to account
  ``associations'' (i.e., characteristics) in acknowledgement that individual
  users may have more than one account and/or that classifications were not
  necessarily mutually exclusive---for example, some ``non-profit generalists''
  had prior lived experience of homelessness.}\BibitemShut {Stop}%
\bibitem [{\citenamefont {Koepfler}\ and\ \citenamefont
  {Hansen}(2012)}]{wearevisibleshort}%
  \BibitemOpen
  \bibfield  {author} {\bibinfo {author} {\bibfnamefont {J.~A.}\ \bibnamefont
  {Koepfler}}\ and\ \bibinfo {author} {\bibfnamefont {D.~L.}\ \bibnamefont
  {Hansen}},\ }\bibfield  {title} {\bibinfo {title} {We {A}re {V}isible:
  Technology-mediated social participation in a {T}witter network for the
  homeless},\ }\href@noop {} {\bibfield  {journal} {\bibinfo  {journal}
  {Proceedings of the 2012 iConference. ACM}\ ,\ \bibinfo {pages} {492}}
  (\bibinfo {year} {2012})}\BibitemShut {NoStop}%
\bibitem [{\citenamefont {Alajajian}\ \emph {et~al.}(2017)\citenamefont
  {Alajajian}, \citenamefont {Williams}, \citenamefont {Reagan}, \citenamefont
  {Alajajian}, \citenamefont {Frank}, \citenamefont {Mitchell}, \citenamefont
  {Lahne}, \citenamefont {Danforth},\ and\ \citenamefont
  {Dodds}}]{lexicocalorimeter}%
  \BibitemOpen
  \bibfield  {author} {\bibinfo {author} {\bibfnamefont {S.}~\bibnamefont
  {Alajajian}}, \bibinfo {author} {\bibfnamefont {J.}~\bibnamefont {Williams}},
  \bibinfo {author} {\bibfnamefont {A.}~\bibnamefont {Reagan}}, \bibinfo
  {author} {\bibfnamefont {S.}~\bibnamefont {Alajajian}}, \bibinfo {author}
  {\bibfnamefont {M.}~\bibnamefont {Frank}}, \bibinfo {author} {\bibfnamefont
  {L.}~\bibnamefont {Mitchell}}, \bibinfo {author} {\bibfnamefont
  {J.}~\bibnamefont {Lahne}}, \bibinfo {author} {\bibfnamefont
  {C.}~\bibnamefont {Danforth}},\ and\ \bibinfo {author} {\bibfnamefont
  {P.}~\bibnamefont {Dodds}},\ }\bibfield  {title} {\bibinfo {title} {The
  {L}exicocalorimeter: {G}auging public health through caloric input and output
  on social media},\ }\href@noop {} {\  (\bibinfo {year} {2017})}\BibitemShut
  {NoStop}%
\bibitem [{\citenamefont {Mitchell}\ \emph {et~al.}(2013)\citenamefont
  {Mitchell}, \citenamefont {Frank}, \citenamefont {Harris}, \citenamefont
  {Dodds},\ and\ \citenamefont {Danforth}}]{happiness}%
  \BibitemOpen
  \bibfield  {author} {\bibinfo {author} {\bibfnamefont {L.}~\bibnamefont
  {Mitchell}}, \bibinfo {author} {\bibfnamefont {M.}~\bibnamefont {Frank}},
  \bibinfo {author} {\bibfnamefont {K.}~\bibnamefont {Harris}}, \bibinfo
  {author} {\bibfnamefont {P.}~\bibnamefont {Dodds}},\ and\ \bibinfo {author}
  {\bibfnamefont {C.}~\bibnamefont {Danforth}},\ }\bibfield  {title} {\bibinfo
  {title} {The {G}eography of {H}appiness: {C}onnecting {T}witter {S}entiment
  and {E}xpression, {D}emographics, and {O}bjective {C}haracteristics of
  {P}lace},\ }\bibfield  {journal} {\bibinfo  {journal} {PLoS ONE}\ }\textbf
  {\bibinfo {volume} {8}},\ \href
  {https://doi.org/https://doi.org/10.1371/journal.pone.0064417}
  {https://doi.org/10.1371/journal.pone.0064417} (\bibinfo {year}
  {2013})\BibitemShut {NoStop}%
\bibitem [{\citenamefont {Ramadona}\ \emph {et~al.}(2016)\citenamefont
  {Ramadona}, \citenamefont {Agusta}, \citenamefont {Sulistyawati},
  \citenamefont {Lazuardi}, \citenamefont {Holmner}, \citenamefont {Dewi},
  \citenamefont {Kusnanto},\ and\ \citenamefont {Röcklov}}]{denguefeverI}%
  \BibitemOpen
  \bibfield  {author} {\bibinfo {author} {\bibfnamefont {A.}~\bibnamefont
  {Ramadona}}, \bibinfo {author} {\bibfnamefont {R.}~\bibnamefont {Agusta}},
  \bibinfo {author} {\bibnamefont {Sulistyawati}}, \bibinfo {author}
  {\bibfnamefont {L.}~\bibnamefont {Lazuardi}}, \bibinfo {author}
  {\bibfnamefont {A.~C.~A.}\ \bibnamefont {Holmner}}, \bibinfo {author}
  {\bibfnamefont {F.}~\bibnamefont {Dewi}}, \bibinfo {author} {\bibfnamefont
  {H.}~\bibnamefont {Kusnanto}},\ and\ \bibinfo {author} {\bibfnamefont
  {J.}~\bibnamefont {Röcklov}},\ }\bibfield  {title} {\bibinfo {title} {Mining
  of health and disease events on {T}witter: Validating search protocols within
  the setting of {I}ndonesia}\ }\href
  {https://doi.org/https://doi.org/10.48550/ARXIV.1608.05910}
  {https://doi.org/10.48550/ARXIV.1608.05910} (\bibinfo {year}
  {2016})\BibitemShut {NoStop}%
\bibitem [{\citenamefont {Ramadona}\ \emph {et~al.}(2019)\citenamefont
  {Ramadona}, \citenamefont {Tozan}, \citenamefont {Lazuardi},\ and\
  \citenamefont {Rocklöv}}]{denguefeverII}%
  \BibitemOpen
  \bibfield  {author} {\bibinfo {author} {\bibfnamefont {A.}~\bibnamefont
  {Ramadona}}, \bibinfo {author} {\bibfnamefont {Y.}~\bibnamefont {Tozan}},
  \bibinfo {author} {\bibfnamefont {L.}~\bibnamefont {Lazuardi}},\ and\
  \bibinfo {author} {\bibfnamefont {J.}~\bibnamefont {Rocklöv}},\ }\bibfield
  {title} {\bibinfo {title} {A combination of incidence data and mobility
  proxies from social media predicts the intra-urban spread of dengue in
  {Y}ogyakarta, {I}ndonesia},\ }\bibfield  {journal} {\bibinfo  {journal} {PLoS
  Neglected Tropical Diseases}\ }\textbf {\bibinfo {volume} {13}},\ \href
  {https://doi.org/https://doi.org/10.1371/journal.pntd.0007298}
  {https://doi.org/10.1371/journal.pntd.0007298} (\bibinfo {year}
  {2019})\BibitemShut {NoStop}%
\bibitem [{Note2()}]{Note2}%
  \BibitemOpen
  \bibinfo {note} {No tweets were returned for the period January 1, 2010,
  through February 28, 2010.}\BibitemShut {Stop}%
\bibitem [{\citenamefont {Alshaabi}\ \emph {et~al.}(2021)\citenamefont
  {Alshaabi}, \citenamefont {Adams}, \citenamefont {Arnold}, \citenamefont
  {Minot}, \citenamefont {Dewhurst}, \citenamefont {Reagan}, \citenamefont
  {Danforth},\ and\ \citenamefont {Dodds}}]{storywrangler}%
  \BibitemOpen
  \bibfield  {author} {\bibinfo {author} {\bibfnamefont {T.}~\bibnamefont
  {Alshaabi}}, \bibinfo {author} {\bibfnamefont {J.~L.}\ \bibnamefont {Adams}},
  \bibinfo {author} {\bibfnamefont {M.~V.}\ \bibnamefont {Arnold}}, \bibinfo
  {author} {\bibfnamefont {J.~R.}\ \bibnamefont {Minot}}, \bibinfo {author}
  {\bibfnamefont {D.~R.}\ \bibnamefont {Dewhurst}}, \bibinfo {author}
  {\bibfnamefont {A.~J.}\ \bibnamefont {Reagan}}, \bibinfo {author}
  {\bibfnamefont {C.~M.}\ \bibnamefont {Danforth}},\ and\ \bibinfo {author}
  {\bibfnamefont {P.~S.}\ \bibnamefont {Dodds}},\ }\bibfield  {title} {\bibinfo
  {title} {Storywrangler: A massive exploratorium for sociolinguistic,
  cultural, socioeconomic, and political timelines using {T}witter},\ }\href
  {https://storywrangling.org/} {\bibfield  {journal} {\bibinfo  {journal}
  {Science Advances}\ }\textbf {\bibinfo {volume} {7}} (\bibinfo {year}
  {2021})}\BibitemShut {NoStop}%
\bibitem [{Note3()}]{Note3}%
  \BibitemOpen
  \bibinfo {note} {Due to many states' departures from previous Point-in-Time
  count practices during the COVID-19 pandemic, the 2020 and 2021 counts were
  excluded from our analysis.}\BibitemShut {Stop}%
\bibitem [{\citenamefont {{United States Census Bureau}}(2023)}]{popestimates}%
  \BibitemOpen
  \bibfield  {author} {\bibinfo {author} {\bibnamefont {{United States Census
  Bureau}}},\ }\href
  {https://www.census.gov/data/developers/data-sets/acs-1year.html} {\bibinfo
  {title} {American {C}ommunity {S}urvey 1-{Y}ear {D}ata (2005-2021) [{D}ata
  set]}} (\bibinfo {year} {2023})\BibitemShut {NoStop}%
\bibitem [{\citenamefont {{United States Census}}(2010)}]{landestimates}%
  \BibitemOpen
  \bibfield  {author} {\bibinfo {author} {\bibnamefont {{United States
  Census}}},\ }\href
  {https://www.census.gov/geographies/reference-files/2010/geo/state-area.html}
  {\bibinfo {title} {State {A}rea {M}easurements and {I}nternal {P}oint
  {C}oordinates [{D}ata set]}} (\bibinfo {year} {2010})\BibitemShut {NoStop}%
\bibitem [{Note4()}]{Note4}%
  \BibitemOpen
  \bibinfo {note} {We did not consider the impact of proxy servers on location
  distribution.}\BibitemShut {Stop}%
\bibitem [{\citenamefont {{geopy}}(2006)}]{geopy}%
  \BibitemOpen
  \bibfield  {author} {\bibinfo {author} {\bibnamefont {{geopy}}},\ }\href
  {https://github.com/geopy/geopy} {\bibinfo {title} {geopy v. 2.3.0 [{C}ode,
  {P}ython]}} (\bibinfo {year} {2006})\BibitemShut {NoStop}%
\bibitem [{\citenamefont {Linnell}\ \emph {et~al.}(2021)\citenamefont
  {Linnell}, \citenamefont {Arnold}, \citenamefont {Alshaabi}, \citenamefont
  {McAndrew}, \citenamefont {Lim}, \citenamefont {Dodds},\ and\ \citenamefont
  {Danforth}}]{sleeploss}%
  \BibitemOpen
  \bibfield  {author} {\bibinfo {author} {\bibfnamefont {K.}~\bibnamefont
  {Linnell}}, \bibinfo {author} {\bibfnamefont {M.}~\bibnamefont {Arnold}},
  \bibinfo {author} {\bibfnamefont {T.}~\bibnamefont {Alshaabi}}, \bibinfo
  {author} {\bibfnamefont {T.}~\bibnamefont {McAndrew}}, \bibinfo {author}
  {\bibfnamefont {J.}~\bibnamefont {Lim}}, \bibinfo {author} {\bibfnamefont
  {P.~S.}\ \bibnamefont {Dodds}},\ and\ \bibinfo {author} {\bibfnamefont
  {C.~M.}\ \bibnamefont {Danforth}},\ }\bibfield  {title} {\bibinfo {title}
  {The sleep loss insult of spring daylight savings in the us is observable in
  twitter activity},\ }\href@noop {} {\bibfield  {journal} {\bibinfo  {journal}
  {Journal of Big Data}\ }\textbf {\bibinfo {volume} {8}} (\bibinfo {year}
  {2021})}\BibitemShut {NoStop}%
\bibitem [{\citenamefont {Dodds}\ \emph {et~al.}(2020)\citenamefont {Dodds},
  \citenamefont {Minot}, \citenamefont {Arnold}, \citenamefont {Alshaabi},
  \citenamefont {Adams}, \citenamefont {Dewhurst}, \citenamefont {Gray},
  \citenamefont {Frank}, \citenamefont {Reagan},\ and\ \citenamefont
  {Danforth}}]{allotaxonometry}%
  \BibitemOpen
  \bibfield  {author} {\bibinfo {author} {\bibfnamefont {P.~S.}\ \bibnamefont
  {Dodds}}, \bibinfo {author} {\bibfnamefont {J.~R.}\ \bibnamefont {Minot}},
  \bibinfo {author} {\bibfnamefont {M.~V.}\ \bibnamefont {Arnold}}, \bibinfo
  {author} {\bibfnamefont {T.}~\bibnamefont {Alshaabi}}, \bibinfo {author}
  {\bibfnamefont {J.~L.}\ \bibnamefont {Adams}}, \bibinfo {author}
  {\bibfnamefont {D.~R.}\ \bibnamefont {Dewhurst}}, \bibinfo {author}
  {\bibfnamefont {T.~J.}\ \bibnamefont {Gray}}, \bibinfo {author}
  {\bibfnamefont {M.~R.}\ \bibnamefont {Frank}}, \bibinfo {author}
  {\bibfnamefont {A.~J.}\ \bibnamefont {Reagan}},\ and\ \bibinfo {author}
  {\bibfnamefont {C.~M.}\ \bibnamefont {Danforth}},\ }\href@noop {} {\bibinfo
  {title} {Allotaxonometry and rank-turbulence divergence: {A} universal
  instrument for comparing complex systems}} (\bibinfo {year} {2020}),\
  \bibinfo {note} {available online at
  \href{https://arxiv.org/abs/2002.09770}{https://arxiv.org/abs/2002.09770}}\BibitemShut
  {NoStop}%
\bibitem [{\citenamefont {Dodds}\ \emph {et~al.}(2011)\citenamefont {Dodds},
  \citenamefont {Harris}, \citenamefont {Kloumann}, \citenamefont {Bliss},\
  and\ \citenamefont {Danforth}}]{hedonometer}%
  \BibitemOpen
  \bibfield  {author} {\bibinfo {author} {\bibfnamefont {P.~S.}\ \bibnamefont
  {Dodds}}, \bibinfo {author} {\bibfnamefont {K.~D.}\ \bibnamefont {Harris}},
  \bibinfo {author} {\bibfnamefont {I.~M.}\ \bibnamefont {Kloumann}}, \bibinfo
  {author} {\bibfnamefont {C.~A.}\ \bibnamefont {Bliss}},\ and\ \bibinfo
  {author} {\bibfnamefont {C.~M.}\ \bibnamefont {Danforth}},\ }\bibfield
  {title} {\bibinfo {title} {Temporal {P}atterns of {H}appiness and
  {I}nformation in a {G}lobal {S}ocial {N}etwork: {H}edonometrics and
  {T}witter},\ }\bibfield  {journal} {\bibinfo  {journal} {PLOS One}\ }\textbf
  {\bibinfo {volume} {6}},\ \href
  {https://doi.org/https://doi.org/10.1371/journal.pone.0026752}
  {https://doi.org/10.1371/journal.pone.0026752} (\bibinfo {year}
  {2011})\BibitemShut {NoStop}%
\bibitem [{\citenamefont {Reagan}\ \emph {et~al.}(2017)\citenamefont {Reagan},
  \citenamefont {Danforth}, \citenamefont {Tivnan}, \citenamefont {Williams},\
  and\ \citenamefont {Dodds}}]{sentiment}%
  \BibitemOpen
  \bibfield  {author} {\bibinfo {author} {\bibfnamefont {A.~J.}\ \bibnamefont
  {Reagan}}, \bibinfo {author} {\bibfnamefont {C.~M.}\ \bibnamefont
  {Danforth}}, \bibinfo {author} {\bibfnamefont {B.}~\bibnamefont {Tivnan}},
  \bibinfo {author} {\bibfnamefont {J.~R.}\ \bibnamefont {Williams}},\ and\
  \bibinfo {author} {\bibfnamefont {P.~S.}\ \bibnamefont {Dodds}},\ }\bibfield
  {title} {\bibinfo {title} {Sentiment analysis methods for understanding
  large-scale texts: a case for using continuum-scored words and word shift
  graphs},\ }\bibfield  {journal} {\bibinfo  {journal} {EPJ Data Science}\
  }\textbf {\bibinfo {volume} {6}},\ \href
  {https://doi.org/https://doi.org/10.1140/epjds/s13688-017-0121-9}
  {https://doi.org/10.1140/epjds/s13688-017-0121-9} (\bibinfo {year}
  {2017})\BibitemShut {NoStop}%
\bibitem [{Note5()}]{Note5}%
  \BibitemOpen
  \bibinfo {note} {Words with neutral scores that add little to semantic
  meaning, such as ``of}\BibitemShut {NoStop}%
\bibitem [{\citenamefont {Dodds}\ \emph {et~al.}(2015)\citenamefont {Dodds},
  \citenamefont {Clark}, \citenamefont {Desu}, \citenamefont {Frank},
  \citenamefont {Reagan}, \citenamefont {Williams}, \citenamefont {Mitchell},
  \citenamefont {Harris}, \citenamefont {Kloumann}, \citenamefont {Bagrow},
  \citenamefont {Megerdoomian}, \citenamefont {McMahon}, \citenamefont
  {Tivnan},\ and\ \citenamefont {Danforth}}]{polyannaprinciple}%
  \BibitemOpen
  \bibfield  {author} {\bibinfo {author} {\bibfnamefont {P.}~\bibnamefont
  {Dodds}}, \bibinfo {author} {\bibfnamefont {E.}~\bibnamefont {Clark}},
  \bibinfo {author} {\bibfnamefont {S.}~\bibnamefont {Desu}}, \bibinfo {author}
  {\bibfnamefont {M.}~\bibnamefont {Frank}}, \bibinfo {author} {\bibfnamefont
  {A.}~\bibnamefont {Reagan}}, \bibinfo {author} {\bibfnamefont
  {J.}~\bibnamefont {Williams}}, \bibinfo {author} {\bibfnamefont
  {L.}~\bibnamefont {Mitchell}}, \bibinfo {author} {\bibfnamefont
  {K.}~\bibnamefont {Harris}}, \bibinfo {author} {\bibfnamefont
  {I.}~\bibnamefont {Kloumann}}, \bibinfo {author} {\bibfnamefont
  {J.}~\bibnamefont {Bagrow}}, \bibinfo {author} {\bibfnamefont
  {K.}~\bibnamefont {Megerdoomian}}, \bibinfo {author} {\bibfnamefont
  {M.}~\bibnamefont {McMahon}}, \bibinfo {author} {\bibfnamefont
  {B.}~\bibnamefont {Tivnan}},\ and\ \bibinfo {author} {\bibfnamefont
  {C.}~\bibnamefont {Danforth}},\ }\bibfield  {title} {\bibinfo {title} {Human
  language reveals a universal positivity bias},\ }\bibfield  {journal}
  {\bibinfo  {journal} {Proceedings of the National Academy of Sciences}\
  }\textbf {\bibinfo {volume} {112}},\ \href
  {https://doi.org/https://doi.org/10.1073/pnas.1411678112}
  {https://doi.org/10.1073/pnas.1411678112} (\bibinfo {year}
  {2015})\BibitemShut {NoStop}%
\bibitem [{Note6()}]{Note6}%
  \BibitemOpen
  \bibinfo {note} {It is noteworthy that sentiment of homelessness tweets tends
  to peak at the end of each year before plummeting at the beginning of the
  next consecutive year. Although the intervals of available homelessness count
  data does not permit comparison at the monthly scale, future research could
  use word shift diagrams to determine whether these cyclical peaks are a
  result of holiday appeals by homeless-serving agencies, followed by concern
  expressed for the homeless during cold-weather months in northern
  states.}\BibitemShut {Stop}%
\bibitem [{\citenamefont {Henry}\ \emph {et~al.}(2017)\citenamefont {Henry},
  \citenamefont {Watt}, \citenamefont {Rosenthal}, \citenamefont {Shivji},\
  and\ \citenamefont {{Abt Associates}}}]{ahar2017}%
  \BibitemOpen
  \bibfield  {author} {\bibinfo {author} {\bibfnamefont {M.}~\bibnamefont
  {Henry}}, \bibinfo {author} {\bibfnamefont {R.}~\bibnamefont {Watt}},
  \bibinfo {author} {\bibfnamefont {L.}~\bibnamefont {Rosenthal}}, \bibinfo
  {author} {\bibfnamefont {A.}~\bibnamefont {Shivji}},\ and\ \bibinfo {author}
  {\bibnamefont {{Abt Associates}}},\ }\href
  {https://www.huduser.gov/portal/sites/default/files/pdf/2017-AHAR-Part-1.pdf}
  {\bibinfo {title} {The {U}.{S}. {D}epartment of {H}ousing and {U}rban
  {D}evelopment, 2017. {T}he 2017 {A}nnual {H}omeless {A}ssessment {R}eport
  (ahar) to {C}ongress}} (\bibinfo {year} {2017})\BibitemShut {NoStop}%
\bibitem [{\citenamefont {Li}\ \emph {et~al.}(2019)\citenamefont {Li},
  \citenamefont {Hu}, \citenamefont {Zhang},\ and\ \citenamefont
  {Zhao}}]{beast}%
  \BibitemOpen
  \bibfield  {author} {\bibinfo {author} {\bibfnamefont {Y.}~\bibnamefont
  {Li}}, \bibinfo {author} {\bibfnamefont {T.}~\bibnamefont {Hu}}, \bibinfo
  {author} {\bibfnamefont {X.}~\bibnamefont {Zhang}},\ and\ \bibinfo {author}
  {\bibfnamefont {K.}~\bibnamefont {Zhao}},\ }\href
  {https://github.com/zhaokg/Rbeast} {\bibinfo {title} {R{B}east: Bayesian
  {C}hange-{P}oint {D}etection and {T}ime {S}eries {D}ecomposition [{C}ode,
  {P}ython]}} (\bibinfo {year} {2019})\BibitemShut {NoStop}%
\bibitem [{Note7()}]{Note7}%
  \BibitemOpen
  \bibinfo {note} {Colorado and New Mexico, negligibly negative scores of -0.01
  and -0.003, respectively}\BibitemShut {NoStop}%
\bibitem [{\citenamefont {{Computational Story Lab}}(2019)}]{allotaxcode}%
  \BibitemOpen
  \bibfield  {author} {\bibinfo {author} {\bibnamefont {{Computational Story
  Lab}}},\ }\href {https://gitlab.com/compstorylab/allotaxonometer} {\bibinfo
  {title} {allotaxonometer [{C}ode, {M}atlab]}} (\bibinfo {year}
  {2019})\BibitemShut {NoStop}%
\bibitem [{\citenamefont {Gallagher}\ \emph {et~al.}(2021)\citenamefont
  {Gallagher}, \citenamefont {Frank}, \citenamefont {Mitchell}, \citenamefont
  {Schwartz}, \citenamefont {Reagan}, \citenamefont {Danforth},\ and\
  \citenamefont {Dodds}}]{wordshifts}%
  \BibitemOpen
  \bibfield  {author} {\bibinfo {author} {\bibfnamefont {R.~J.}\ \bibnamefont
  {Gallagher}}, \bibinfo {author} {\bibfnamefont {M.~R.}\ \bibnamefont
  {Frank}}, \bibinfo {author} {\bibfnamefont {L.}~\bibnamefont {Mitchell}},
  \bibinfo {author} {\bibfnamefont {A.~J.}\ \bibnamefont {Schwartz}}, \bibinfo
  {author} {\bibfnamefont {A.~J.}\ \bibnamefont {Reagan}}, \bibinfo {author}
  {\bibfnamefont {C.~M.}\ \bibnamefont {Danforth}},\ and\ \bibinfo {author}
  {\bibfnamefont {P.~S.}\ \bibnamefont {Dodds}},\ }\bibfield  {title} {\bibinfo
  {title} {Generalized {W}ord {S}hift {G}raphs: A {M}ethod for {V}isualizing
  and {E}xplaining {P}airwise {C}omparisons between texts},\ }\href@noop {}
  {\bibfield  {journal} {\bibinfo  {journal} {EPJ Data Science}\ }\textbf
  {\bibinfo {volume} {10}} (\bibinfo {year} {2021})}\BibitemShut {NoStop}%
\bibitem [{Note8()}]{Note8}%
  \BibitemOpen
  \bibinfo {note} {We included 2016 in the latter period because the first PiT
  count to register a nationwide increase took place in January 2017, meaning
  that the increase took place retroactively during 2016}\BibitemShut {NoStop}%
\bibitem [{Note9()}]{Note9}%
  \BibitemOpen
  \bibinfo {note} {E.g., ``daca'', ``registerhomeless'',
  ``pelosi''}\BibitemShut {NoStop}%
\bibitem [{Note10()}]{Note10}%
  \BibitemOpen
  \bibinfo {note} {The emphasis on immigration-related words further suggests
  that the issue of homelessness began to be contextualized within broader
  political pressure to respond to a reported influx of undocumented
  immigration to the US via the US-Mexico border, covered at length by American
  news media starting in 2018 at the end of ex-President Trump's term in
  office.}\BibitemShut {Stop}%
\bibitem [{Note11()}]{Note11}%
  \BibitemOpen
  \bibinfo {note} {Note that Top 20 Individual and Top 20 Entity accounts are
  all multiple-post accounts}\BibitemShut {NoStop}%
\bibitem [{\citenamefont {{United States Department of Housing and Urban
  Development}}(2014)}]{pitmethodguide}%
  \BibitemOpen
  \bibfield  {author} {\bibinfo {author} {\bibnamefont {{United States
  Department of Housing and Urban Development}}},\ }\bibfield  {title}
  {\bibinfo {title} {H{U}{D} {P}oint-in-{T}ime {C}ount {M}ethodology {G}uide},\
  }\href
  {https://files.hudexchange.info/resources/documents/PIT-Count-Methodology-Guide.pdf}
  {\  (\bibinfo {year} {2014})}\BibitemShut {NoStop}%
\bibitem [{Note12()}]{Note12}%
  \BibitemOpen
  \bibinfo {note} {The percent of household income spent, on average, for
  housing costs}\BibitemShut {NoStop}%
\end{thebibliography}%

\clearpage

\onecolumngrid
 
\appendix

\setcounter{page}{1}
\renewcommand{\thepage}{A\arabic{page}}
\renewcommand{\thefigure}{A\arabic{figure}}
\renewcommand{\thetable}{A\arabic{table}}
\setcounter{figure}{0}
\setcounter{table}{0}
 
\section{Allotaxonometry plots}
\label{sec:measuring-homelessness.allotax}

\begin{figure}[hp!]
    \centering
    \includegraphics[width=\textwidth]{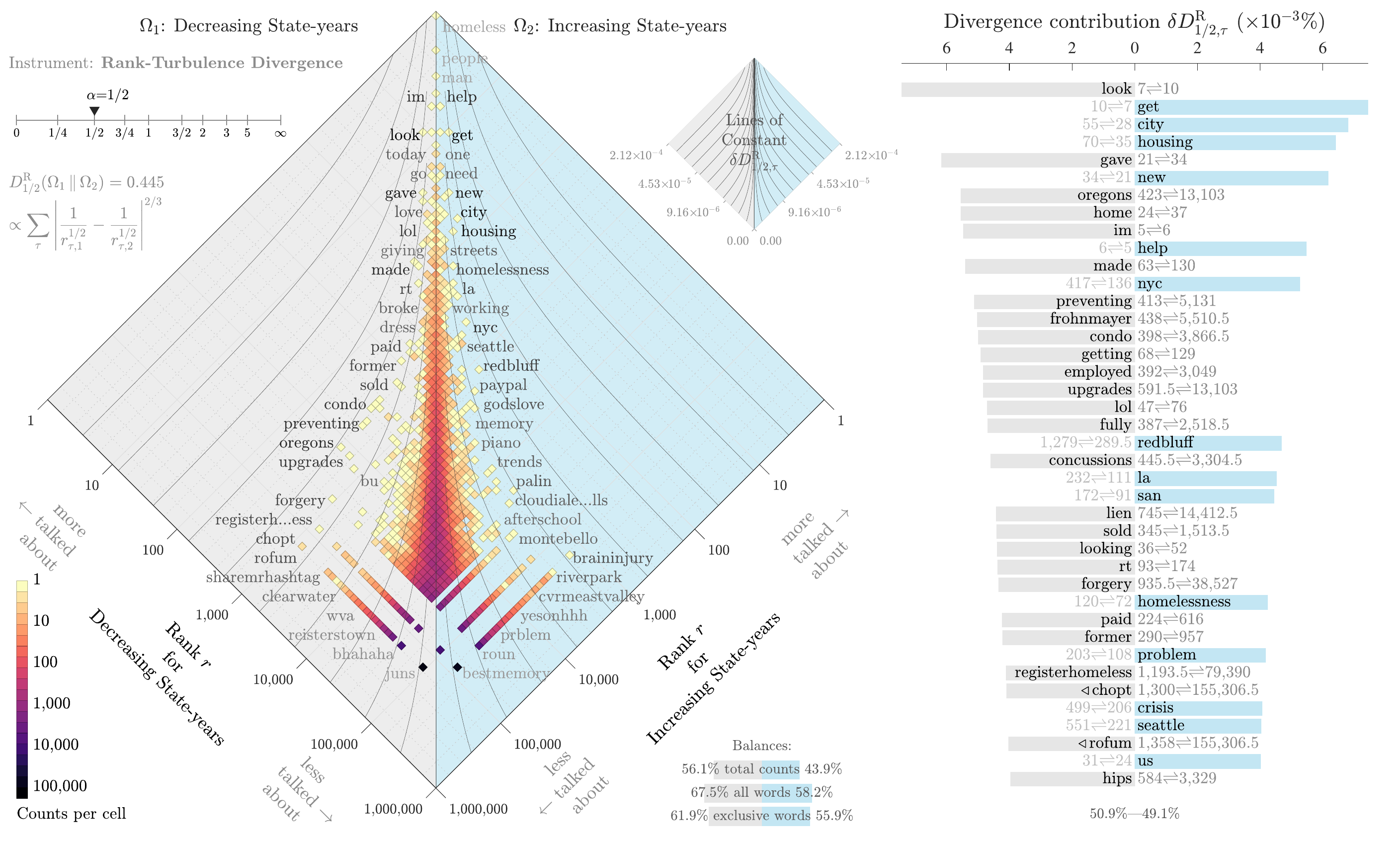}
    \caption{
        \textbf{Comparisons of tweets from state-years with increasing versus decreasing rates of homelessness.}
            }
    \label{Fig:allotaxstateyears}
\end{figure}

\begin{figure}[hp!]
\centering
    \includegraphics[width=\textwidth]{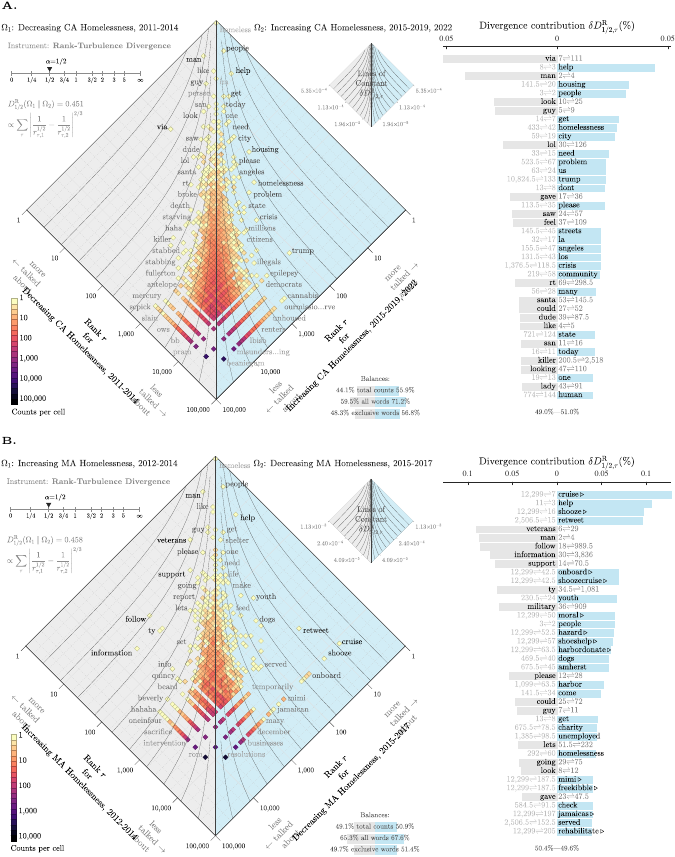}
\caption{   
\textbf{Alloxonographs for tweets from \textbf{A.} California and \textbf{B.} Massachusetts, comparing pre- and post-polarity shift in homelessness count trend.} 
For California, the earlier years are characterized by increasing homelessness, 
which is represented on the left-hand side of the histogram, while for Massachusetts, the reverse is true.
}
    \label{Fig:allotaxCAMA}
\end{figure}

\begin{figure}
    \centering
    \includegraphics[width=\textwidth]{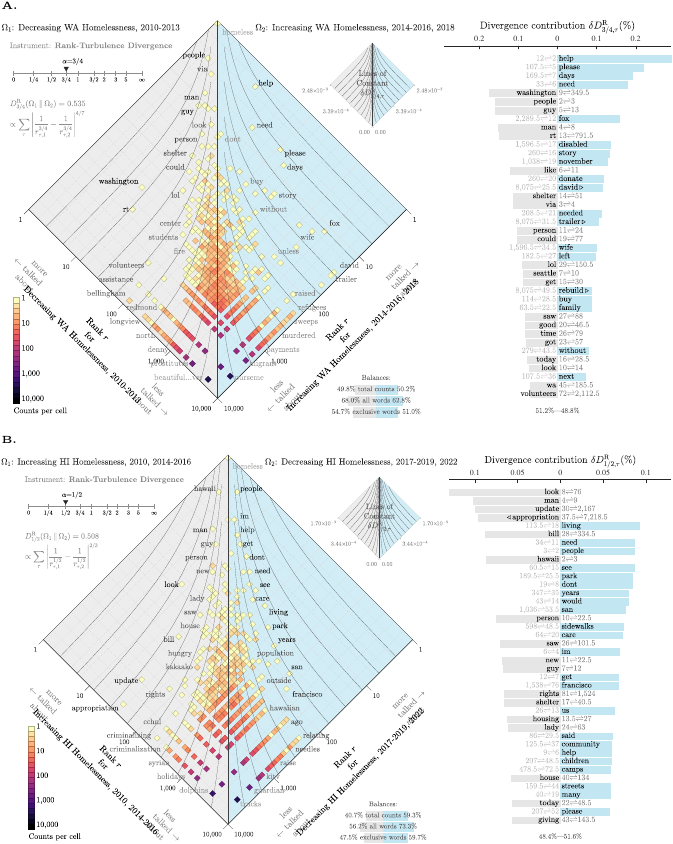}
    \label{Fig:allotaxHIWA}    
    \caption{
        \textbf{Allotaxonographs for tweets from \textbf{A.} Washington and \textbf{B.} Hawaii compared to pre- and post-polarity shift in homelessness count trend.} 
        As for California in Fig.~\ref{Fig:allotaxCAMA}, the earlier years for Washington reflect an increase in homelessness, while Hawaii is similar to Massachusetts.
        }
\end{figure}

\clearpage

\section{Storywrangler plots}
\label{sec:measuring-homelessness.storywrangler}

\mbox{}

\begin{figure}[hp!]
    \centering
    \includegraphics[width=\textwidth]{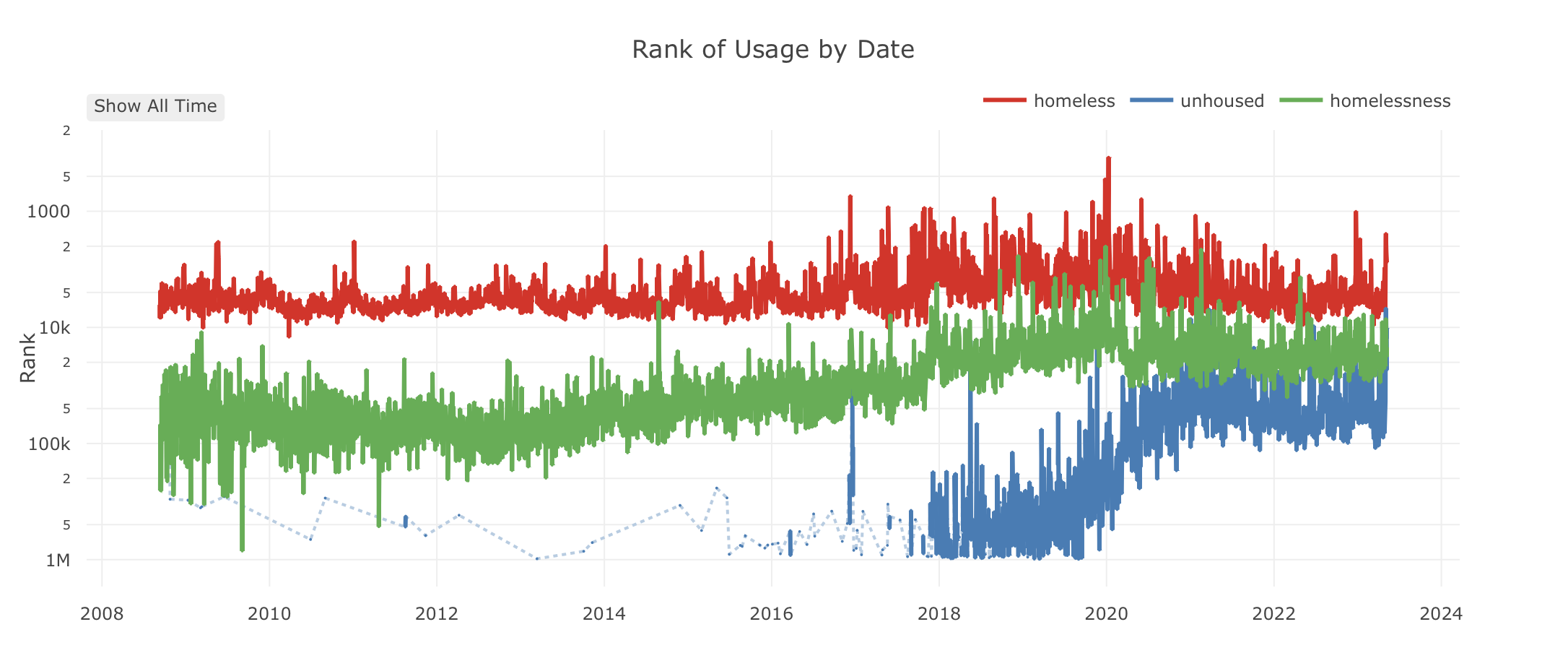}
    \label{Fig:unhoused}    
    \caption{
        \textbf{Log-odds for 1-grams ``homeless'', ``homelessness'', and ``unhoused'' from 2010--2023.} Although not the subject of our research, other 1-grams may also reference homelessness. ``Unhoused'' in particular gains momentum in late 2017.
        }
\end{figure}

\begin{figure}[tp!]
    \centering
    \includegraphics[width=0.75\textwidth]{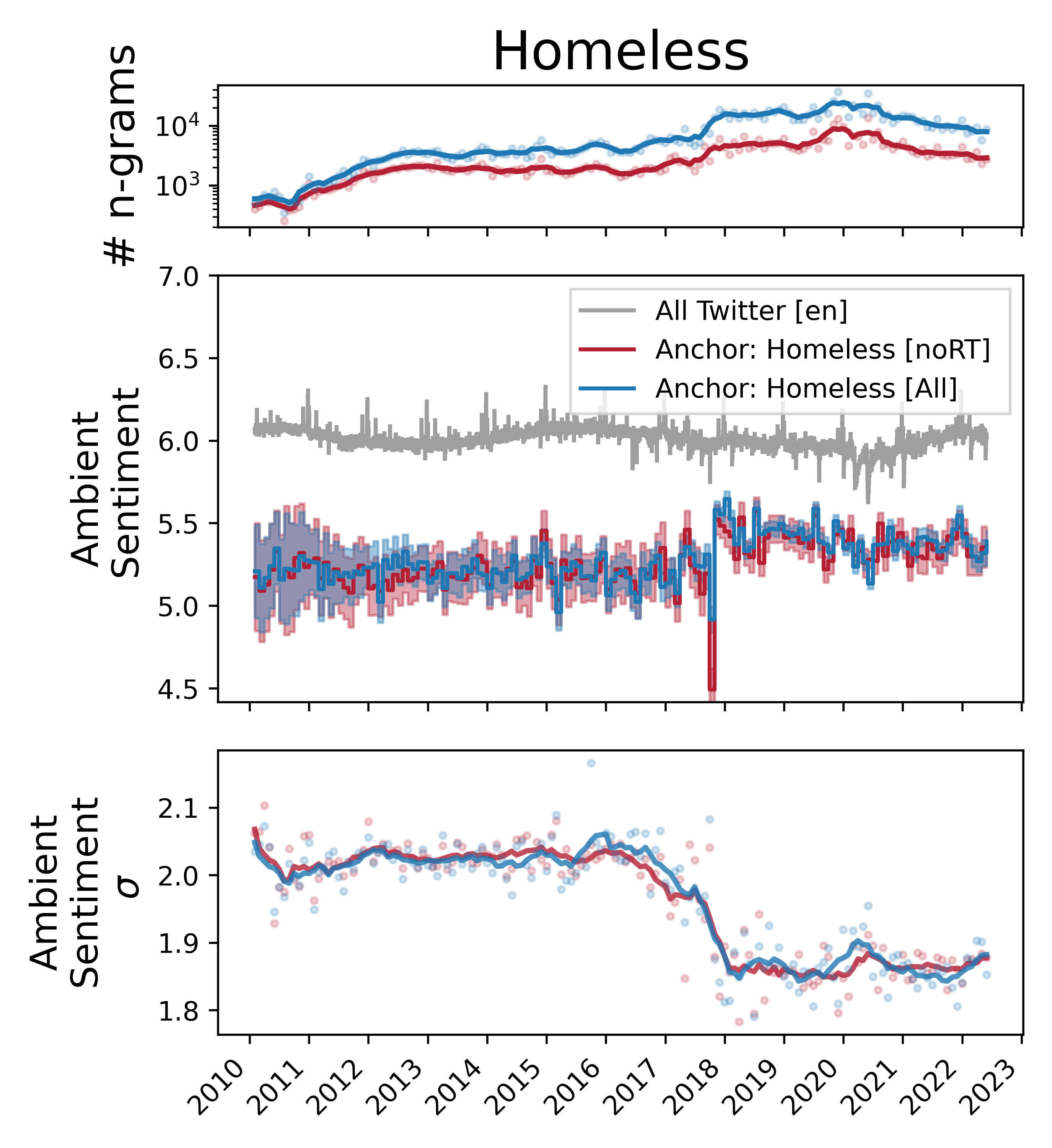}    

    \caption{
    \textbf{Ambient sentiment of a 10\% sample of all English-language tweets containing the word ``homeless'', as compared with ambient sentiment of all of English-language Twitter in the same period.} The blue line represents only original tweets, while the red line includes both original tweets and all retweets. Ambient sentiment changes present in US-geotagged tweets are reflected in the broader English-language population and are not unique to geotagged tweets.}
    \label{fig:all_eng_sentiment}
\end{figure}

\clearpage

\section{Comparison of high-homelessness versus low-homelessness states}
\label{sec:measuring-homelessness.high-low-comparison}

\begin{table}[hp!]
\begin{tabular}{|c|c|c||c|c|c|}
         \hline
      Rank 
      & 
      High Homelessness 
      & 
      Low Homelessness 
      & 
      Rank 
      & 
      High Homelessness 
      & 
      Low Homelessness \\
    \hline
    \rowcolor{Gray} 1  &  \cellcolor{blue!25}housing  &  much  &  26  &  covid19my  &  would  \\
2  &  man  &  please  &  27  &  \cellcolor{blue!25}condo  &  lake  \\
\rowcolor{Gray} 3  &  people  &  im  &  28  &  \cellcolor{blue!25}frohnmayer  &  want  \\
4  &  \cellcolor{blue!25}homelessness  &  look  &  29  &  thousands  &  make  \\
\rowcolor{Gray} 5  &  street  &  trying  &  30  &  fully  &  think  \\
6  &  w  &  venmo  &  31  &  \cellcolor{blue!25}preventing  &  \cellcolor{blue!25}vets  \\
\rowcolor{Gray} 7  &  train  &  elliaivy  &  32  & \cellcolor{blue!25} employed  &  could  \\
8  &  new  &  patrickemcashapp  &  33  &  former  &  care  \\
\rowcolor{Gray} 9  &  \cellcolor{blue!25}county  &  really  &  34  &  streets  &  going  \\
10  &  \cellcolor{blue!25}youth  &  raise  &  35  &  g  &  money  \\
\rowcolor{Gray} 11  &  \cellcolor{blue!25}crisis  &  \cellcolor{blue!25}transgender  &  36  &  sidewalk  &  today  \\
12  &  \cellcolor{blue!25}encampment  &  possible  &  37  &  years  &  person  \\
\rowcolor{Gray} 13  &  woman  &  need  &  38  &  ppl  &  dress  \\
14  &  made  &  thank  &  39  &  support  &  crowdfund  \\
\rowcolor{Gray} 15  &  wheelchair  &  share  &  40  &  1  &  \cellcolor{blue!25}home  \\
16  &  via  &  donate  &  41  &  real  &  discipled  \\
\rowcolor{Gray} 17  &  us  &  paypal  &  42  &  beach  &  \cellcolor{blue!25}house  \\
18  &  videoemail  &  lgbtqia  &  43  &  human  &  downtown  \\
\rowcolor{Gray} 19  &  ignored  &  help  &  44  &  paid  &  converted  \\
20  &  subway  &  like  &  45  &  link  &  starving  \\
\rowcolor{Gray} 21  &  video  &  shelter  &  46  &  problem  &  mexico  \\
22  &  user  &  dont  &  47  &  santa  &  get  \\
\rowcolor{Gray} 23  &  getting  &  looking  &  48  &  black  &  thats  \\
24  &  sold  &  go  &  49  &  station  &  cute  \\
\rowcolor{Gray} 25  &  4  &  salt  &  50  &  \cellcolor{blue!25}city  &  many  \\
    \hline
\end{tabular}
\caption{\textbf{The top 50 words in each corpus contributing to differences between tweets from consistently high-homelessness states versus tweets from low-homelessness states.} Rankings correspond to the size of the difference of representation in either corpus.}
\end{table}

\begin{table}

\begin{tabular}{|c|c|c||c|c|c|}
         \hline
    Rank & High Homelessness & Low Homelessness & Rank & High Homelessness & Low Homelessness \\
    \hline
    \rowcolor{Gray} 1  &  \cellcolor{blue!25}homeless  &  cruise  &  26  &  last  &  across  \\
    2  &  \cellcolor{blue!25}veterans  &  help  &  27  &  night  &  meals  \\
    \rowcolor{Gray} 3  &  man  &  shooze  &  28  &  think  &  \cellcolor{blue!25}americans  \\
    4  &  follow  &  retweet  &  29  &  feel  &  started  \\
    \rowcolor{Gray} 5  &  look  &  \cellcolor{blue!25}youth  &  30  &  got  &  usa  \\
    6  &  guy  &  onboard  &  31  &  right  &  plus  \\
    \rowcolor{Gray} 7  &  im  &  shoozecruise  &  32  &  dont  &  15m  \\
    8  &  like  &  moral  &  33  &  opened  &  jamaica  \\
    \rowcolor{Gray} 9  &  support  &  hazard  &  34  &  us  &  freekibble  \\
    10  &  please  &  amherst  &  35  &  set  &  11yo  \\
    \rowcolor{Gray} 11  &  information  &  shoeshelp  &  36  &  drunk  &  mimi  \\
    12  &  ty  &  dogs  &  37  &  info  &  jamaicas  \\
    \rowcolor{Gray} 13  &  \cellcolor{blue!25}military  &  harbordonate  &  38  &  iphune  &  better  \\
    14  &  person  &  harbor  &  39  &  next  &  rehabilitate  \\
    \rowcolor{Gray} 15  &  could  &  get  &  40  &  dude  &  w  \\
    16  &  lets  &  come  &  41  &  back  &  local  \\
    \rowcolor{Gray} 17  &  going  &  aug12  &  42  &  change  &  make  \\
    18  &  people  &  charity  &  43  &  goal  &  u  \\
    \rowcolor{Gray} 19  &  lol  &  4  &  44  &  school  &  cats  \\
    20  &  gave  &  \cellcolor{blue!25}homelessness  &  45  &  2  &  young  \\
    \rowcolor{Gray} 21  &  report  &  \cellcolor{blue!25}unemployed  &  46  &  looking  &  warm  \\
    22  &  gonna  &  check  &  47  &  fire  &  \cellcolor{blue!25}housing  \\
    \rowcolor{Gray} 23  &  bad  &  served  &  48  &  dress  &  \cellcolor{blue!25}vets  \\
    24  &  today  &  \cellcolor{blue!25}shelter  &  49  &  park  &  land  \\
    \rowcolor{Gray} 25  &  told  &  feed  &  50  &  shit  &  common  \\
      \hline
\end{tabular}
\caption{\textbf{Words that are more common in tweets generated during Massachusetts's period of increasing (2012-2014) versus decreasing (2015-2017) homelessness, ranked by the size of the difference of representation in either corpus.}}
\end{table}

\clearpage

\section{Notation and measurement glossary}
\label{sec:measuring-homelessness.glossary}

\begin{table*}[h!]

    \begin{tcolorbox}[width=\linewidth,colback={white},title={\textbf{Homelessness volume, quantified}},colbacktitle=Gray,coltitle=black]    
 
        \begin{tabular}{l|l}
             Per capita total homelessness count & $HC_s^{(y)} = \frac{\text{year y homeless count in state s}}{\text{total state s population in year y}}$  \\
             &\\
             Per square mile homelessness density & $ HD_s^{(y)} = \frac{\text{year y homeless count in state s}}{\text{total square land-miles in state s}}$\\
             &\\
             General population density & $ GD_s^{(y)} = \frac{\text{year y total state s population }}{\text{total square land-miles in state s}}$\\
             &\\
             Per capita homelessness-related tweet count & $ HT_s^{(y)} = \frac{\text{year y homelessness tweet count state s}}{\text{total state s population in year y}}$ \\
             &\\
             Log-odds ratio of homelessness-related tweets & $\log_{10}\Big{(}{\frac{\text{count of English-lang 1-grams ``homeless''}}{\text{total count of English-lang 1-grams}}}\Big{)}$\\
             &\\
             Annualized rate of change, homelessness count & $HC_s^{(y)}-HC_s^{(y-1)}$\\
             &\\
             Annualized rate of change, homelessness density & $HD_s^{(y)}-HD_s^{(y-1)}$\\
        \end{tabular}

    \end{tcolorbox} 

\label{fig:measuring-homelessness.glossary}
\end{table*}


\end{document}